\documentclass[a4paper,fleqn]{cas-sc}

\usepackage{float}
\usepackage{algorithm}
\usepackage{algpseudocode}
\usepackage{tikz}
\usepackage{graphicx}
\usepackage{subcaption}
\usetikzlibrary{calc}
\usepackage{stfloats}

\usepackage[numbers]{natbib}
\def\tsc#1{\csdef{#1}{\textsc{\lowercase{#1}}\xspace}}
\tsc{WGM}
\tsc{QE}

\newcommand{\brac}[1]  {\left( #1 \right)}

\usepackage{caption}
\usepackage[font=small,labelfont=bf]{caption}

\newif\ifanonymous
\anonymousfalse      % arXiv version

\begin{document}
\let\WriteBookmarks\relax
\def\floatpagepagefraction{1}
\def\textpagefraction{.001}
\newcommand{\etal}[0]{\emph{et al.}}
\shorttitle{Generative Diffusion Model for 3D Turbomachinery Design}    

\ifanonymous
    \shortauthors{Author}
\else
    \shortauthors{Y. Geng et~al.}
\fi

\title [mode = title]{A New Paradigm for 3D Turbomachinery Design: Generative Diffusion Model Based Framework with Direct Geometry Encoding}  




\ifanonymous
    \author[1]{Anonymous Author}
    \affiliation[1]{Anonymous Affiliation}

\else
    \author[1]{Yingfan Geng}[orcid=0009-0004-1175-6302]
    
    \credit{Writing – original draft, Data Curation, Methodology, Software, Formal analysis, Investigation}
    
    \author[1]{Jinhong Wang}[orcid=0000-0002-0012-7218]
    
    \credit{Writing – original draft, Supervision, Conceptualization, Methodology, Formal analysis}
    
    \author[1]{Lazaros Papachristodoulou}[orcid=0009-0007-1911-1631]
    \credit{Writing – original draft, Software, Data Curation, Validation}
    
    \author[2]{Sibo Cheng}[orcid=0000-0002-8707-2589]
    \credit{Writing – review \& editing draft, Formal analysis}
    
    \author[1]{Teng Cao}[orcid=0009-0000-2594-9157] % confirm?
    \credit{Writing – review \& editing draft, Supervision, Conceptualization, Formal analysis}
    \cormark[1]
    
    \affiliation[1]{organization={Department of Mechanical Engineering, Imperial College London},
                addressline={South Kensington Campus}, 
                city={London},
                postcode={SW7 2AZ}, 
                state={London},
                country={United Kingdom}}
    \affiliation[2]{organization={CEREA, Ecole des Ponts and EDF R\&D, Institut Polytechnique de Paris},
                state={Île-de-France},
                country={France}}
    
    \cortext[1]{Corresponding author: t.cao@imperial.ac.uk}
\fi
%  ======================================================

\begin{abstract}
The aerodynamic design of turbomachinery is critical to the performance of the overall energy system, yet it is challenging due to the complex non-linear flow physics and the presence of multiple-target design compromises. Denoising diffusion model, as one of the leading approaches in generative machine learning, has shown its advantages of high design solution accuracy and diversity in many engineering applications. In this study, we bring it to the 3D inverse design problem of turbomachinery, using centrifugal compressors as a classic representative, to demonstrate the new methodology for complex geometry designs. A diffusion model-centred design framework has been developed in this study. By specifying the desired design condition (mass flow rate and rotational speed) and targeted performance (pressure ratio and efficiency), the trained diffusion model returns directly the 3D compressor geometry that satisfies the condition inputs. Compared to traditional deterministic forward design approaches, the proposed method not only generates accurate geometry solutions to inverse design problems, but also enables effective exploration of the entire design space, providing a diverse set of candidate solutions. In addition, this paper presents the first study to directly train on 3D blade geometry coordinates rather than parametrised representations, demonstrating the feasibility of coordinate-based learning while enabling a highly flexible framework applicable to a wide range of designs. The trained diffusion model achieves excellent design capability, with solution accuracy up to 99\% and unfeasible designs less than 1\%. Furthermore, the solution diversity of the trained diffusion model is also quantitatively verified by means of comparing the distribution of the solution sets generated from the diffusion model and from direct sampling of physical parameters. By leveraging diffusion models, the turbomachinery design process is greatly simplified, allowing for the efficient generation of accurate and diverse solutions, thereby marking a milestone in the evolution of turbomachinery design.
\end{abstract}

% \includegraphics[width=\textwidth]{figs/work_flow_chart.pdf}

% \begin{highlights}
%     \item A conditional diffusion model-centered design framework is developed and applied to centrifugal compressor design, demonstrating its capability on turbomachinery designs with complex geometries. 
%     \item The diffusion model is trained on both parametrised and explicit 3D geometries, showing comparable performance. 
%     \item The diffusion model design performance is thoroughly assessed from both the solution accuracy and diversity perspectives. 
% \end{highlights}

%\nocite{*}

% Keywords

\maketitle

\section{Introduction}
\label{section:introduction}

The aerodynamic design for turbomachines is a critical engineering problem, as it significantly affects the overall performance of the energy system. However, the complex 3D blade geometries introduce highly intricate internal flow physics, making the performance sensitive to their aerodynamic designs. Therefore, the design process is a challenging problem, often involving multi-objective optimisation under geometrical constraints~\cite{Came1998}.

With recent developments in machine learning techniques, AI-assisted design and optimisation procedures have emerged in the field of turbomachinery~\cite{Zou2024}. These approaches can generally be divided into two categories: the surrogate model-based forward design loop and the generative model-based inverse design. For forward designs, surrogate models, which are often trained neural network regressors, are used to predict the performance of the turbomachinery designs. The surrogate models aim to replace the CFD simulations or experiments to reduce the costs and time required in the design-validation loops~\cite{Cutrinavilalta2020, Zhang2022}. As for inverse design approaches, generative models take a number of design conditions as inputs and directly return the turbomachinery blade geometry that satisfies the specified conditions. \citet{Jia2025} developed a design framework for 2D compressor blade profile generation, which combines both the machine learning-based forward and backward design approaches. The blade profile geometry is first generated from a conditional invertible neural network, which takes the pressure ratio, pressure losses, and pressure coefficient distribution as inputs. The generated blade profile is then validated using a regression model that is trained from CFD simulation results. This approach is also adopted by \citet{Ghosh2021} on axial turbine inverse designs.

As the probabilistic diffusion model becomes one of the state-of-the-art generative models in machine learning~\cite{Ho2020}, its advantages of generating accurate and diverse design solutions have been proven in multiple engineering disciplines, such as aerofoil designs~\cite{Geng2026} and ship hull designs~\cite{Bagazinski2023}. In addition, these deep learning-based models are auto-differentiable, making them favoured as part of the optimisation method. However, there is only limited research on their applicability in the field of turbomachinery design. \citet{Chen2026} developed a quasi-3D axial compressor blade generation workflow with a diffusion model, where a series of span-wise blade sections are generated, stacked and mapped to form the final 3D design. Diffusion model-based inverse design frameworks were also reported for axial compressor \cite{Wu2025Comp} and turbine blade designs \cite{Wu2025Turb}. This literature shows that by inputting the desired design target (e.g., mass flow rate, efficiency, and pressure ratio), the diffusion model will return a parametrised geometry that can satisfy this condition.

Although diffusion models have demonstrated their potential in inverse design problems for turbomachinery, the limited number of trials generally rely on the geometry parametrisations. For example, the models proposed by \citet{Jia2025} and \citet{Ghosh2021} are conditioned on the pressure coefficient distribution on the blade, which relies on empirical guidelines to generate a sensible distribution. In addition, the majority of the literature trains the generative model using parametrised geometries~\cite{Jia2025, Ghosh2021, Wu2025Comp, Wu2025Turb}. Although geometry parametrisation leads to smaller models and simplifies the training process, it requires prior knowledge (e.g., choice of parameters and curve fittings). It should be noted that the generative models, such as diffusion models, are often originally developed for computer vision tasks~\cite{Ho2020,Karras2022}. They are born to handle data in high-dimensional mathematical spaces, and their full potential in inverse engineering design applications can only be unlocked when they are directly trained with high-dimensional 3D geometries. 

Following the work on 2D aerofoil designs in~\cite{Geng2026}, this paper introduces a diffusion model-based 3D turbomachinery inverse design method that is directly trained on 3D geometries without parameterisation. The centrifugal compressor is selected as the design problem as a representation of the complicated blade geometries in turbomachinery, fully demonstrating the capability of the diffusion model's learning capability on complex 3D geometries and aerodynamic performance. As a baseline for comparison, we also trained another diffusion model using traditional 1D parametrised geometries within the same method. By thoroughly examining the two models' design capabilities from multiple perspectives, we reveal the potential of the diffusion model architecture on high-dimensional geometry data, showing a new paradigm for diffusion model-powered turbomachinery inverse designs. The major contributions of this study are as follows:

\begin{itemize}

    \item The diffusion model is implemented as a novel paradigm of turbomachinery design, and is applied to the centrifugal compressor design as a demonstration. By adopting the state-of-the-art Elucidated Diffusion Models, we delivered an effective, unified inverse design framework to generate centrifugal compressor geometries that satisfy the given design targets ($\eta$ and $\textrm{PR}$ at a given operating $\dot{m}$ and $\omega$) accurately. 
    
    \item To the authors' knowledge, this paper presents the first diffusion model for turbomachinery design that is directly trained on the 3D point coordinates, using a novel and flexible dual-diffusion model architecture. The entire design process can be either constrained with manually specified compressor geometries or be completely automatic with the help of an auxiliary model to explore the geometries. 
    
    \item The 3D diffusion models' design capability is systematically assessed and compared to a baseline 1D model that is trained on 1D parametrised geometries, from both the design accuracy and diversity perspectives. The results consistently show that both the 1D- and 3D-based diffusion models can achieve excellent performance. This proves the concept of applying the diffusion model for non-parametric geometry data, paving the way towards more complicated turbomachinery geometry designs and advanced optimisation methods.

    \item To best utilise the stochastic sampling capability of the diffusion model, a multi-target design procedure is demonstrated. The various design solutions obtained from the diffusion model can all satisfy the primary design targets (PR and $\eta$) with acceptable tolerance, and these solutions can be further selected according to any arbitrary secondary targets, without the need to retrain the diffusion model. 

\end{itemize}

The remainder of the paper is organised as follows. Section~\ref{section:methodology} contains the methodology used in this study, including the compressor geometric and performance data curation, diffusion model implementation, and the model's performance evaluation. The corresponding results are detailed in Section~\ref{section:results} followed by conclusions in Section~\ref{section:conclusion}.

\section{Methodology}
\label{section:methodology}
\subsection{Proposed Design Framework}

\begin{figure*}
    \centering
    \includegraphics[width=\textwidth]{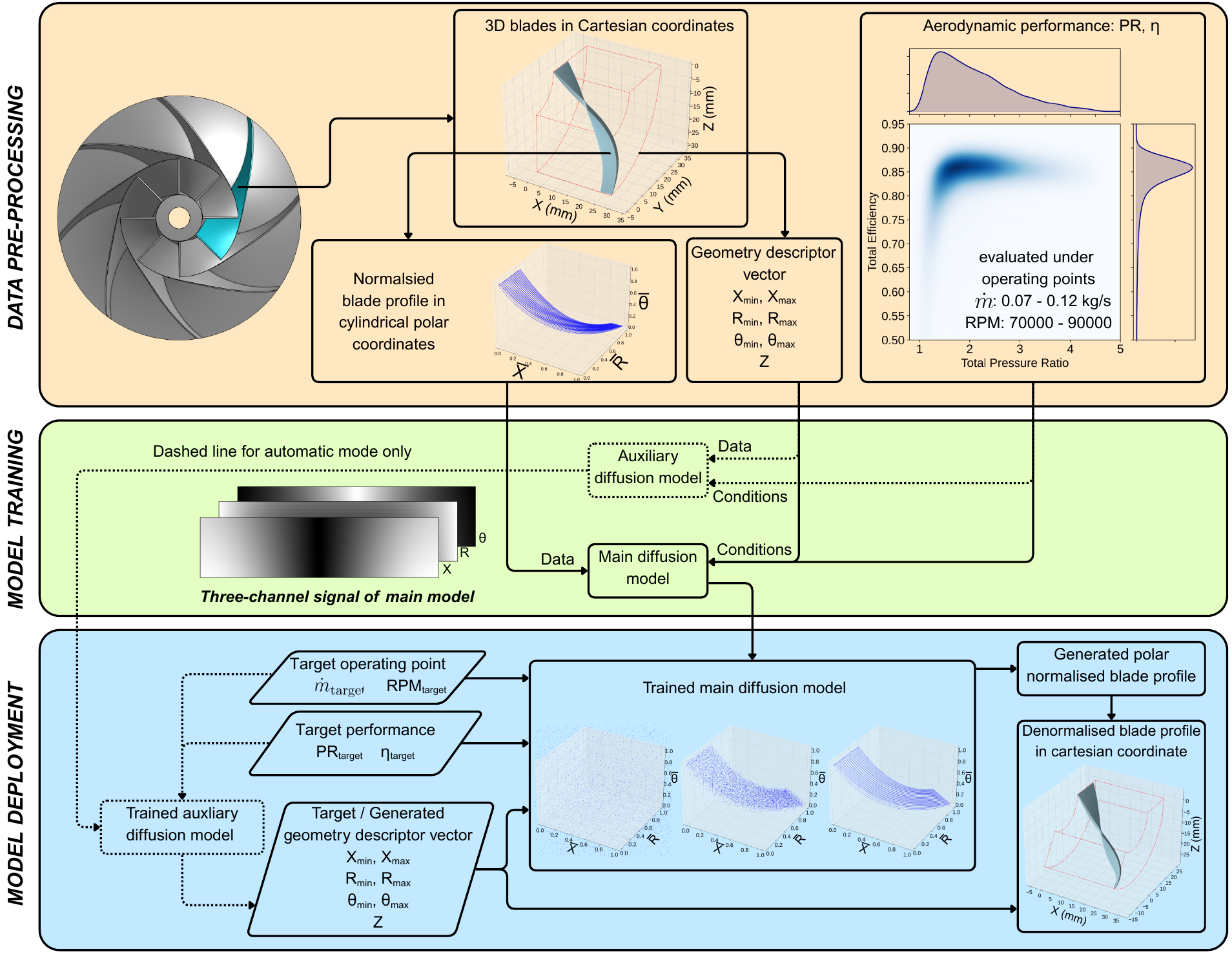}
    \caption{Proposed Centrifugal Compressor Inverse Design Framework with Diffusion Model}
    \label{fig:work_flow_chart}
\end{figure*}

The overall design framework of this study is illustrated by Fig.~\ref{fig:work_flow_chart}. The data curation, model training, and model deployment are explained in detail in Sections~\ref{subsection:data_curation_and_processing}, \ref{subsection:model_implementation}, and \ref{subsection:model_deployment}, respectively.

\subsection{Data Curation \& Preprocessing}
\label{subsection:data_curation_and_processing}
\subsubsection{Geometry Database Generation}
\label{subsubsection:geometry_sampling_and_generation}

\begin{figure}
    \centering
    \includegraphics[width=0.5\columnwidth]{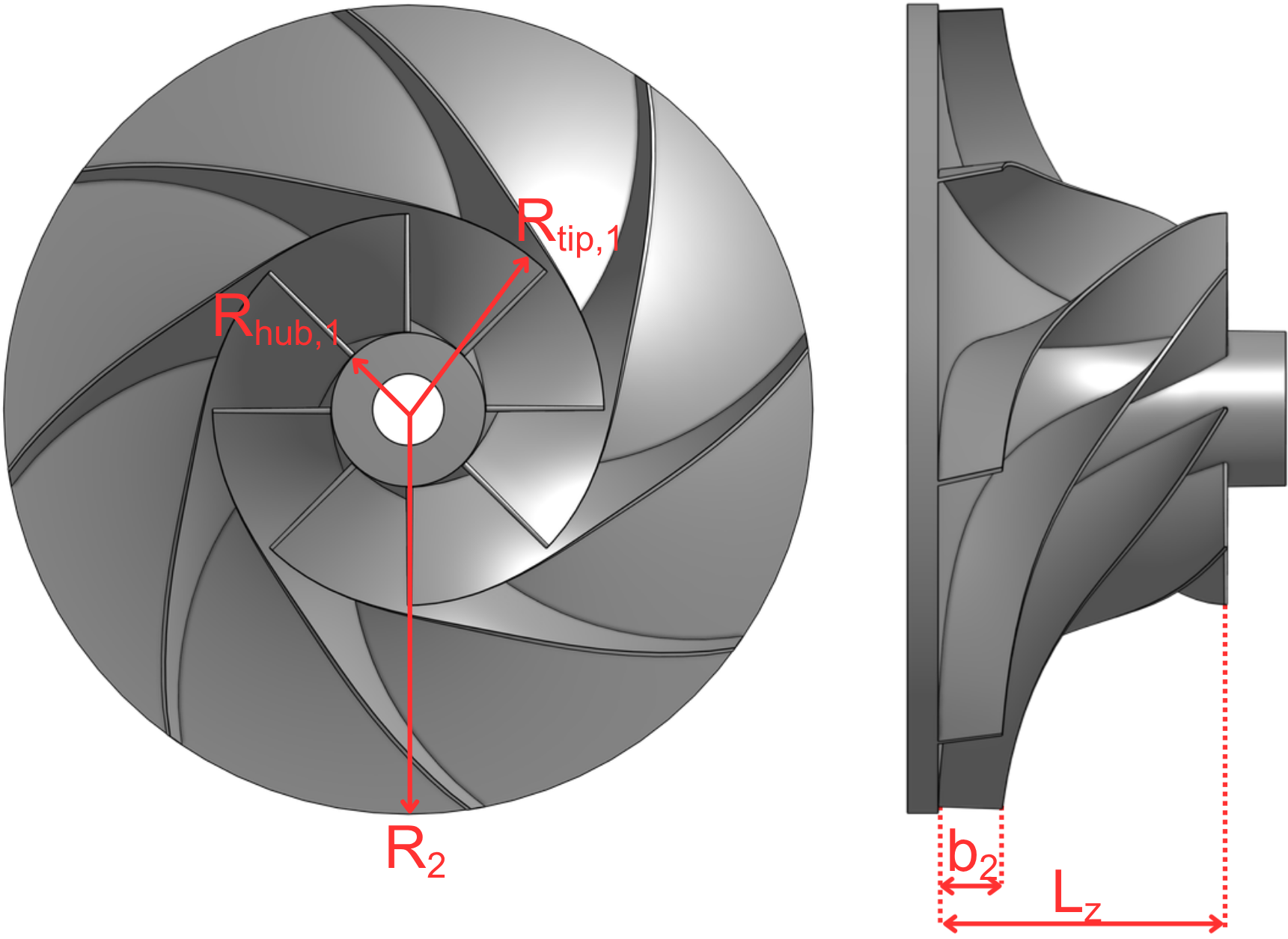}
    \caption{Main Impeller Geometry}
    \label{fig:main_impeller_geometry}
\end{figure}

The first step in this study is to develop a dataset for model training. This consists of two steps, namely the compressor geometry data generation and the corresponding performance data generation, as illustrated by the data pre-processing block in Fig.~\ref{fig:work_flow_chart}. 
% In this study, the compressors consist of an impeller and a vaneless diffuser to form an indicative stage design. 
For demonstration purposes, eight initial design variables are selected and randomly sampled within a typical design range for a subsonic impeller to generate a diverse geometry dataset of centrifugal compressor blade. These variables and their corresponding ranges are summarised in Table~\ref{table:geometry_sampling_variables}. Six out of these eight variables characterise the basic geometry of the compressor impeller, which includes the inlet and outlet radii, blade height, impeller axial length, and the number of impeller blades, as illustrated in Fig.~\ref{fig:main_impeller_geometry}. The remaining two variables (mass flow rate and rotational speed) define the design operating point. In addition, it is assumed that all the compressors operate at standard inlet total conditions ($P_{01} = 101325 ~\textrm{Pa}$, $T_{01} = 288 ~\textrm{K}$) with ideal gas air as the working fluid. The sampling procedure uses the Latin Hypercube Sampling (LHS) algorithm~\cite{Mckay2000}, which is a commonly adopted random sampling methodology in high-dimensional variable spaces. In total, 2500 initial design variable combinations are sampled from the range shown in Table~\ref{table:geometry_sampling_variables}. 

\begin{table}
    \centering
    \renewcommand{\arraystretch}{1.1}
    \caption{Initial design variables and typical design ranges}
    \label{table:geometry_sampling_variables}
    
    \begin{tabular}{lcc}
    \hline
    \textbf{Variable} & \textbf{Symbol} & \textbf{Sampling Range} \\
    \hline
    Inlet Tip Radius       & $R_{\text{tip},1}$  & [10, 30] mm \\
    Inlet Hub Radius       & $R_{\text{hub},1}$  & [7.5, 12.5] mm \\
    Outlet Mean Radius     & $R_{\text{mean},2}$ & [25, 49] mm \\
    Outlet Blade Height    & $b_2$               & [2.9, 6.5] mm \\
    Axial Length           & $L_z$               & [15, 36] mm \\
    Number of Blades       & $Z$                 & [7, 8]  \\
    Design Mass Flow Rate  & $\dot{m}_\text{design}$ & [0.07, 0.12] kg/s \\
    Design Rotational Speed & $\omega_\text{design}$ & [70000, 90000] RPM \\
    \hline
    \end{tabular}
\end{table}

\begin{table}
    \centering
    \caption{The 1D parametrised compressor geometry}
    \label{table:1D_geometry_parameters}
    
    \begin{tabular}{lcc}
        \hline
        \textbf{Description} & \textbf{Symbol} & \textbf{Unit} \\
        \hline
        \multicolumn{3}{c}{\textit{\textbf{Impeller Geometry}}} \\
        \hline
        Inlet tip radius & $R_{\text{tip},1}$ & mm \\
        Inlet hub radius & $R_{\text{hub},1}$ & mm \\
        Inlet blade angle at hub & $\beta_{b1,\text{hub}}$ & deg \\
        Inlet blade angle at tip & $\beta_{b1,\text{tip}}$ & deg \\
        Inlet blade angle at mean radius&$\beta_{b1,\text{mean}}$ & deg \\
        Outlet radius & $R_{2}$ & mm \\
        Outlet blade angle & $\beta_{b2}$ & deg \\
        Outlet blade height & $b_2$ & mm \\
        Axial length & $L_z$ & mm \\
        Tip clearance & $s$ & mm \\
        Blade thickness & $t$ & mm \\
        Number of blades & $Z$ & -- \\
        
        % \hline
        % \multicolumn{3}{c}{\textit{\textbf{Vaneless Diffuser Geometry}}} \\
        % \hline
        % Outlet radius & $R_3$ & mm \\
        % Outlet width & $b_3$ & mm \\
        \hline
    \end{tabular}
\end{table}

The compressor geometry parameters are then derived from the 8 sampled initial design variables, and the complete list is displayed in Table~\ref{table:1D_geometry_parameters}. The 1D parametrised geometry database generation is hence complete and ready to be used for 3D geometry
generation.

An in-house blade forming tool is used to generate the 3D compressor blade geometry from the 1D parameters listed in Table~\ref{table:1D_geometry_parameters}.The tool constructs three independent geometrical distributions: meridional profile, blade angle distribution, and blade thickness distribution. These are then converted into 3D coordinate points, defining the hub, shroud, and blade surfaces. The resulting geometry is stored as point coordinates in the Cartesian coordinate system ($x,y,z$). Each compressor blade consists of 16 span-wise profiles, and on each profile the points are structured, starting from the trailing edge, going to the leading edge along the suction side (SS) before returning to the trailing edge along the pressure side (PS). Each span-wise profile is discretised into 512 coordinate points, resulting in a three-dimensional geometric representation of size $3\times16\times512$ for each blade. At this stage, the 3D compressor geometry database generation is complete. For certain geometric parameter combinations, the blade forming process does not produce a valid geometry. Such cases are flagged as geometrically infeasible and excluded from the dataset. In total, there are approximately 2000 distinct 3D compressor geometries in the database.

\subsubsection{Performance Database Generation}
\label{section: Performance Database Generation}

For demonstration and proof of concept purposes, the compressor performance of each geometry is predicted using a meanline model~\cite{Papachristodoulou2026}. The meanline model is a reduced-order centrifugal compressor performance prediction tool that provides rapid performance evaluations while maintaining reasonable representation of compressor performance. It solves a series of thermodynamic equations and velocity triangles using iterative schemes to determine flow quantities downstream along the meridional path. The irreversibility (aerodynamic losses) of the compressor is determined through a series of empirical loss models accounting major loss mechanism such as incidence, leakage and clearance flow, shock, choke, and friction losses. Although the loss models are empirical in nature, they have been calibrated and validated for centrifugal compressors of comparable scale and operating regimes to those considered in this work~\cite{Papachristodoulou2026}. The present study adopts a simplified full-blade impeller configuration for proof-of-concept demonstration; therefore, the meanline model is used primarily to provide computationally efficient performance estimates suitable for assessing the capabilities of the proposed diffusion-based inverse design framework.

\begin{figure}
    \centering
    \includegraphics[width=0.5\columnwidth]{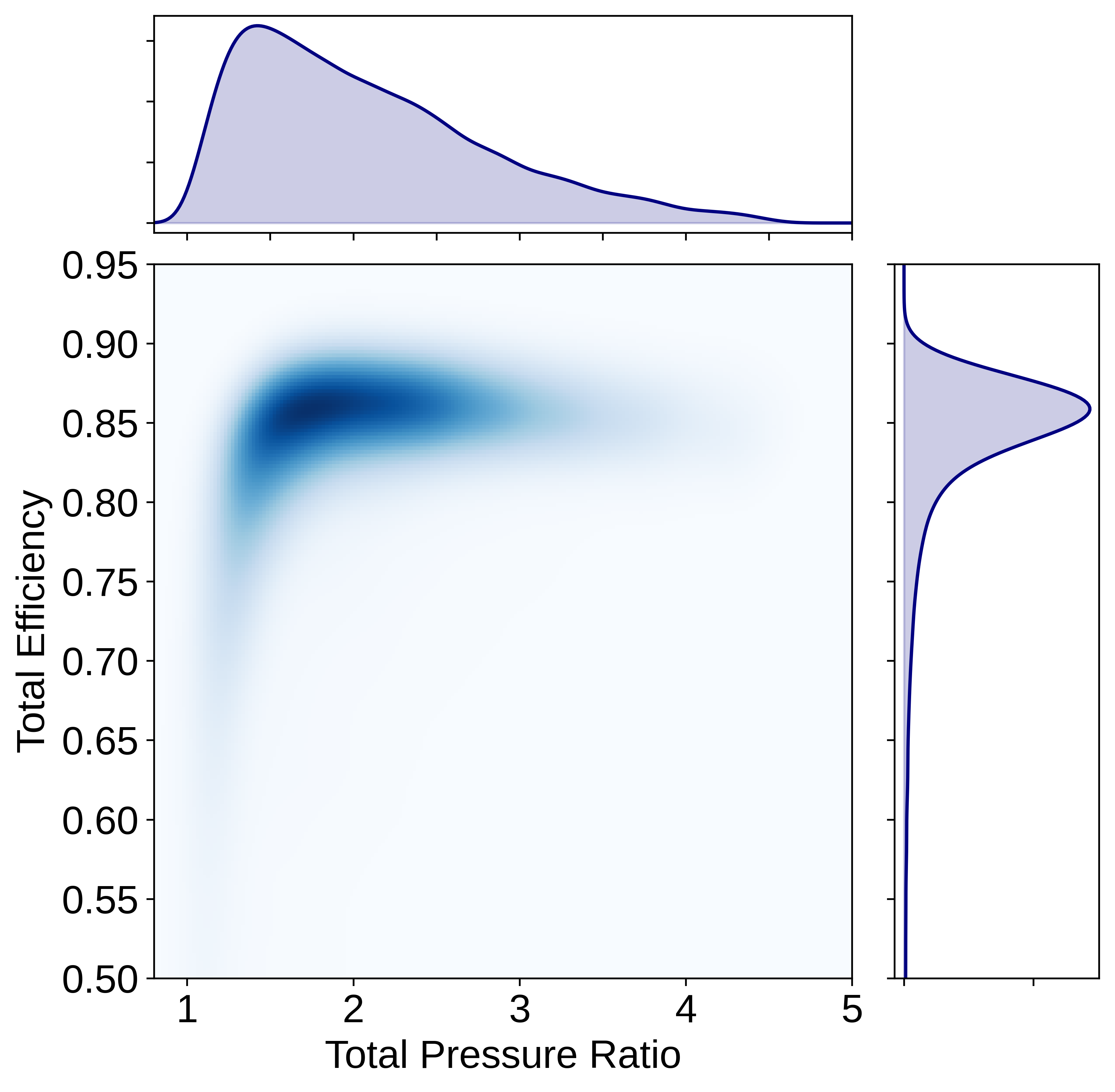}
    \caption{Compressor Performance Database Distribution}
    \label{fig:compressor_performance_distribution}
\end{figure}

All the 1D geometries generated in Section~\ref{subsubsection:geometry_sampling_and_generation} are then fed into the meanline model, and for each of the geometries, the compressor performance is computed at a uniformly sampled operating condition. To be more specific, the mass flow rate ($\dot{m}$) varies in the range of $[0.07,0.12] ~\textrm{kg\,s}^{-1}$ with an increment of 0.01 $\textrm{kg}\,\textrm{s}^{-1}$, and the rotational speed ($\omega$) varies in the range of $[70000, 90000]~\textrm{RPM}$ with an increment of 5000 RPM. In this study, the compressor performance is characterised by the total pressure ratio ($\textrm{PR}$) and total isentropic efficiency $\eta$. It is noted that some operating conditions may be unfeasible for certain geometries, such as when the mass flow rate is too high or too low, causing choking or stalling and resulting in meanline model convergence failure. These non-converging cases are removed from the dataset. The final generated performance database is illustrated by a kernel density estimate (KDE) plot with distributions of $\textrm{PR}$ and $\eta$ in Fig.~\ref{fig:compressor_performance_distribution}. The mathematical formulation of KDE is shown in Eq.~\eqref{eq:kde}.
\begin{equation}
    \label{eq:kde}
    \hat{p}(x)=\frac{1}{n h^d}\sum_{i=1}^{n} \textrm{K} \!\left(\frac{x-x_i}{h}\right)
\end{equation}
where $n$ is the number of data points, $h$ is the bandwidth, $d = 2$ as 2-dimensional data, and $\textrm{K}(\cdot)$ is the kernel function.

\subsubsection{Data Normalisation}
\label{subsubsection:data_normalisation}

\begin{figure*}
    \centering
    \begin{subfigure}{0.32\textwidth}
        \centering
        \includegraphics[width=\textwidth]{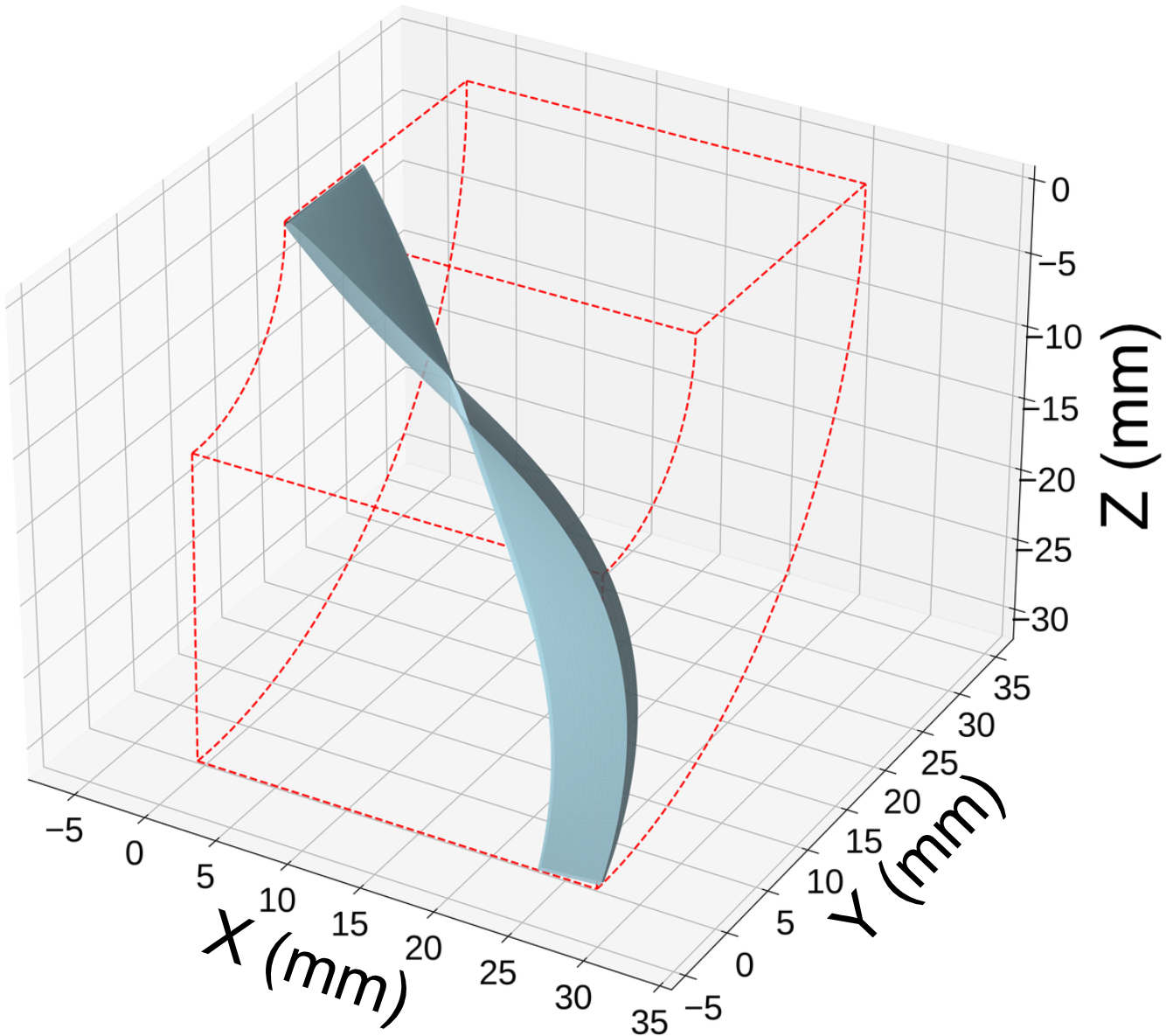}
        \caption{Cartesian $(x,y,z)$}
        \label{subfig:3D_blade_in_cartesian_with_bounding_box}
    \end{subfigure}
    \begin{subfigure}{0.32\textwidth}
        \centering
        \includegraphics[width=\textwidth]{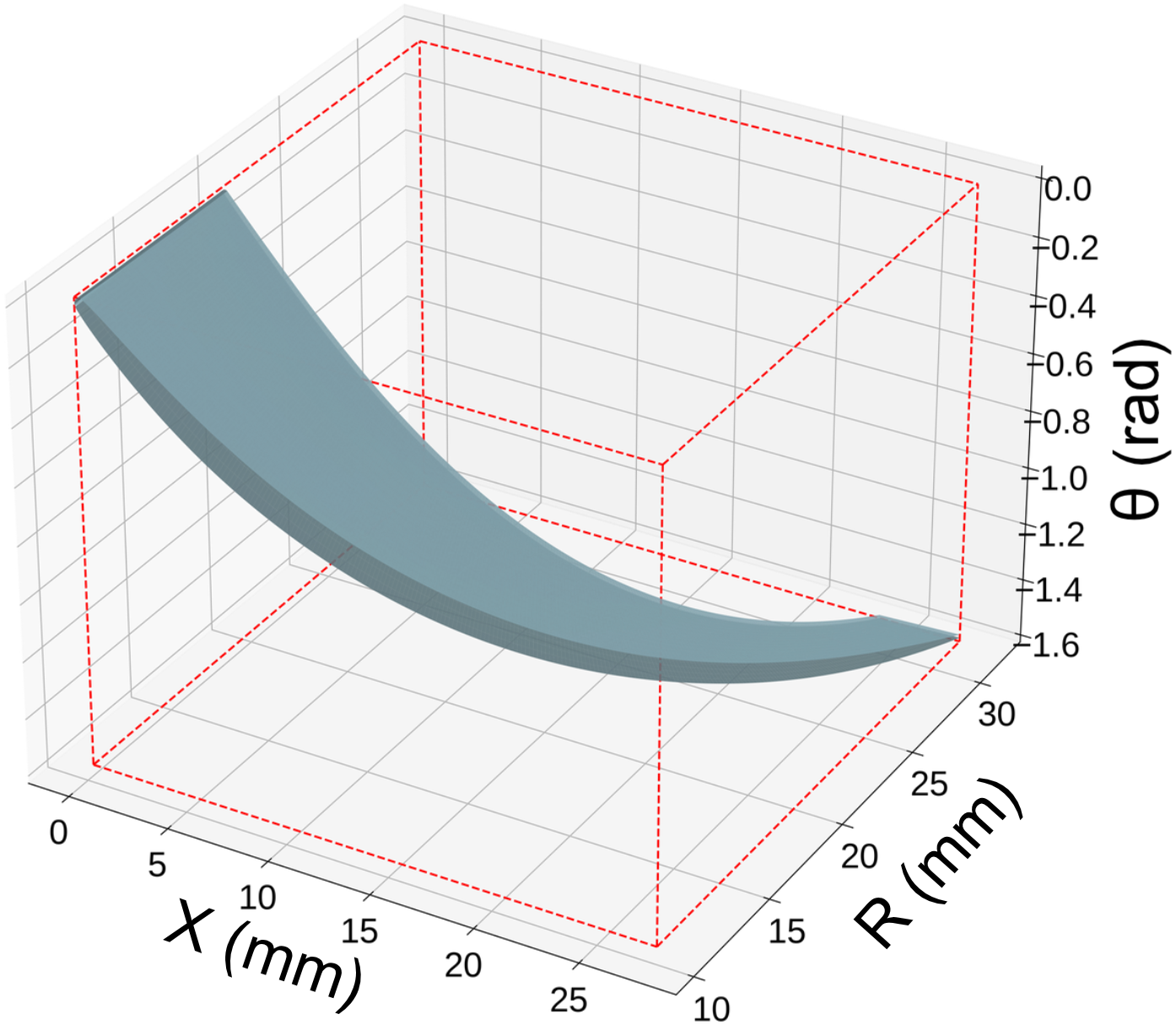}
        \caption{Cylindrical polar $(x,r,\theta)$}
        \label{subfig:compressor_blade_in_polar_coordinates_with_bounding_box}
        \end{subfigure}
    \begin{subfigure}{0.30\textwidth}
    \centering
        \includegraphics[width=\textwidth]{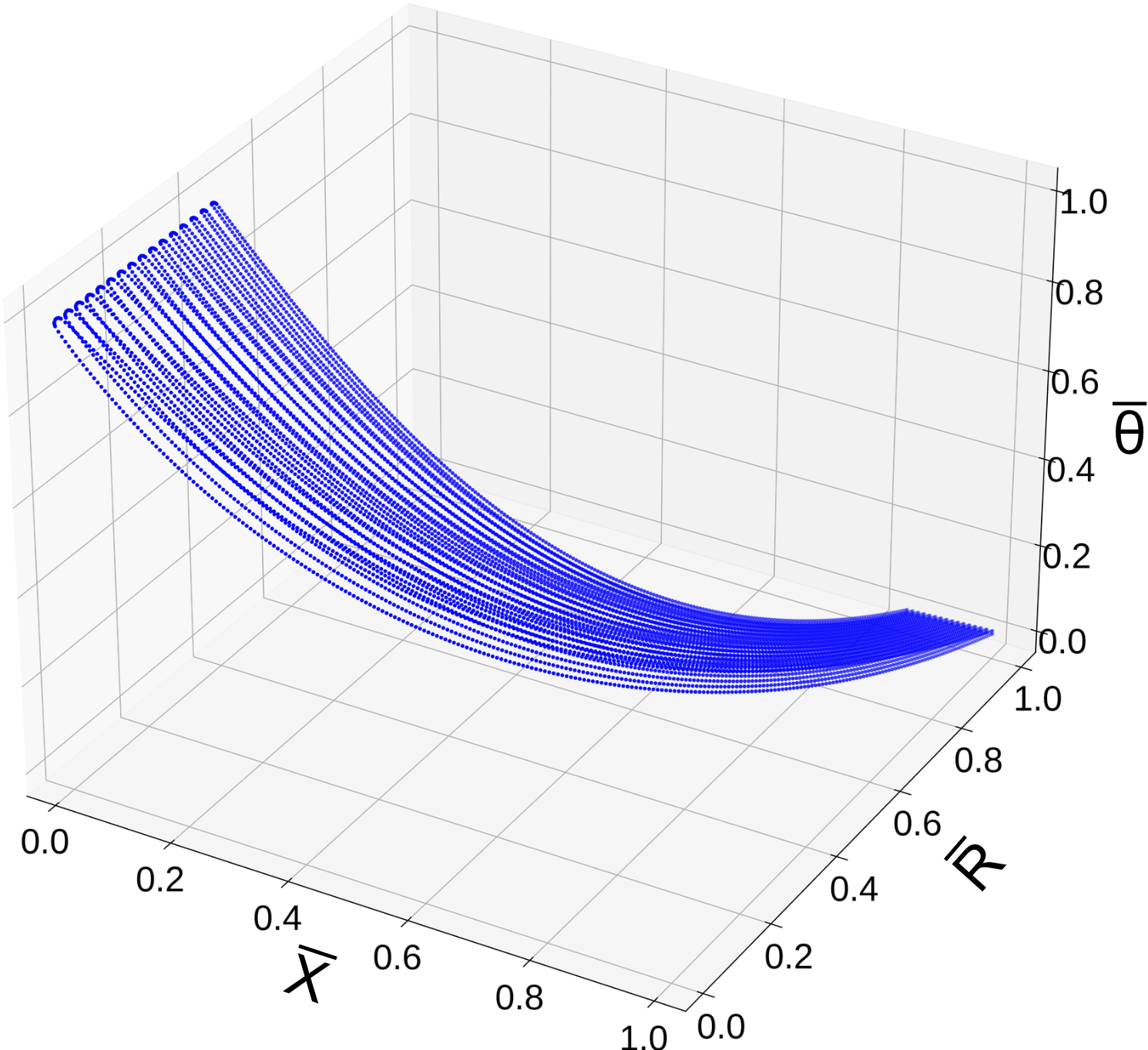}
        \caption{Normalised cylindrical polar $(\overline{x},\overline{r},\overline{\theta})$}
        \label{subfig:cylindrical_polar_coordinate_normalsied_blade}
    \end{subfigure}
    \caption{3D compressor blade geometry presentation in different coordinate systems. To make them visually distinguishable, the blade in the normalised coordinate is shown with discretised points in blue.}
    \label{fig:blade_geometry_normalisation}
\end{figure*}

As a standard practice in deep learning models, all data were normalised to the same range to ensure equal weightings in the model, preventing the large magnitude data from dominating the training process. In this study, both the compressor geometry and the performance labels were scaled to the range of $[0-1]$. 

The global min-max normalisation is shown as Eq.~\eqref{eq:normalisation}, where $\overline{\chi}$ is the normalised value for an arbitrary variable $\chi$ with subscripts min and max denoting the global minimum and maximum values in the entire dataset. 
\begin{equation}
    \label{eq:normalisation}
    \overline{\chi}=\frac{\chi - \chi_{\text{min}}}{\chi_{\text{max}}-\chi_{\text{min}}}
\end{equation}
In this study, min-max normalisation is used for the 1D compressor geometry parameters (Table~\ref{table:1D_geometry_parameters}), operating conditions ($\dot{m}$ and $\omega$), and the performance data ($\textrm{PR}$ and $\eta$).

As for 3D compressor geometry normalisation shown in Fig.~\ref{fig:blade_geometry_normalisation}, additional pre-processing is carried out for two reasons: meaningful geometry presentation and suitability of direct global min-max normalisation. 

% The compressor geometry normalisation procedure is more complicated, with two main concerns. 

Firstly, although the 3D compressor geometries are stored as points in Cartesian coordinates ($x,y,z$), the ranges of these coordinates do not directly reflect meaningful geometric characteristics.Instead, it is more sensible to convert $x,y,z$ coordinates into cylindrical polar coordinates ($x,r,\theta$) as it better aligns with the intrinsic geometric and rotational periodic features of the configuration.

Figure~\ref{subfig:3D_blade_in_cartesian_with_bounding_box} presents a compressor blade in $x,y,z$ coordinates for reference. It is clear that the blades are bounded by a cylindrical sector, as shown by red dashed lines in Fig.~\ref{subfig:3D_blade_in_cartesian_with_bounding_box}. This sector can be presented in a cylindrical polar coordinate system constructed by preserving the original $x$-axis in the Cartesian system as the rotational axis $x$ and $r$, $\theta$ calculated using Eqs.~\eqref{eq:cartesian_to_polar_r} and~\eqref{eq:cartesian_to_polar_theta}, respectively.
\begin{equation}
    \label{eq:cartesian_to_polar_r}
    r = \sqrt{y^2+z^2}
\end{equation}
\begin{equation}
    \label{eq:cartesian_to_polar_theta}
    % \theta = \tan^{-1}\left(\frac{z}{y}\right)
    \theta = \textrm{atan2}(z, y)
\end{equation}
The six faces of the cylindrical sector corresponds to the corresponding range in ($x,r,\theta$), namely $x_{\text{min}}$, $x_{\text{max}}$, $r_{\text{min}}$, $r_{\text{max}}$, $\theta_{\text{min}}$ and $\theta_{\text{max}}$. In this cylindrical polar geometry presentation, it is clear that $x_{\text{max}}-x_{\text{min}}$ is the compressor axial length $L_z$ in Table~\ref{table:1D_geometry_parameters}. The radial range of the compressor blade $r_{\text{min}}$ and $r_{\text{max}}$ are equal to the inlet hub radius $R_{\text{tip},1}$ and outlet radius $R_{2}$, respectively. As for the $\theta$ range, the blade leading edge characterises the $\theta_{\text{min}}$ and the $\theta_{\text{max}}$ is the resulting $\theta$ at the trailing edge hub point. 
After coordinate system conversion, the blade geometry in the cylindrical polar coordinate system ($x,r,\theta$) is illustrated by Fig.~\ref{subfig:compressor_blade_in_polar_coordinates_with_bounding_box}, and the bounding box is now converted to a cuboid in cylindrical polar coordinates. 
\begin{figure}
    \centering
    \includegraphics[width=0.5\columnwidth]{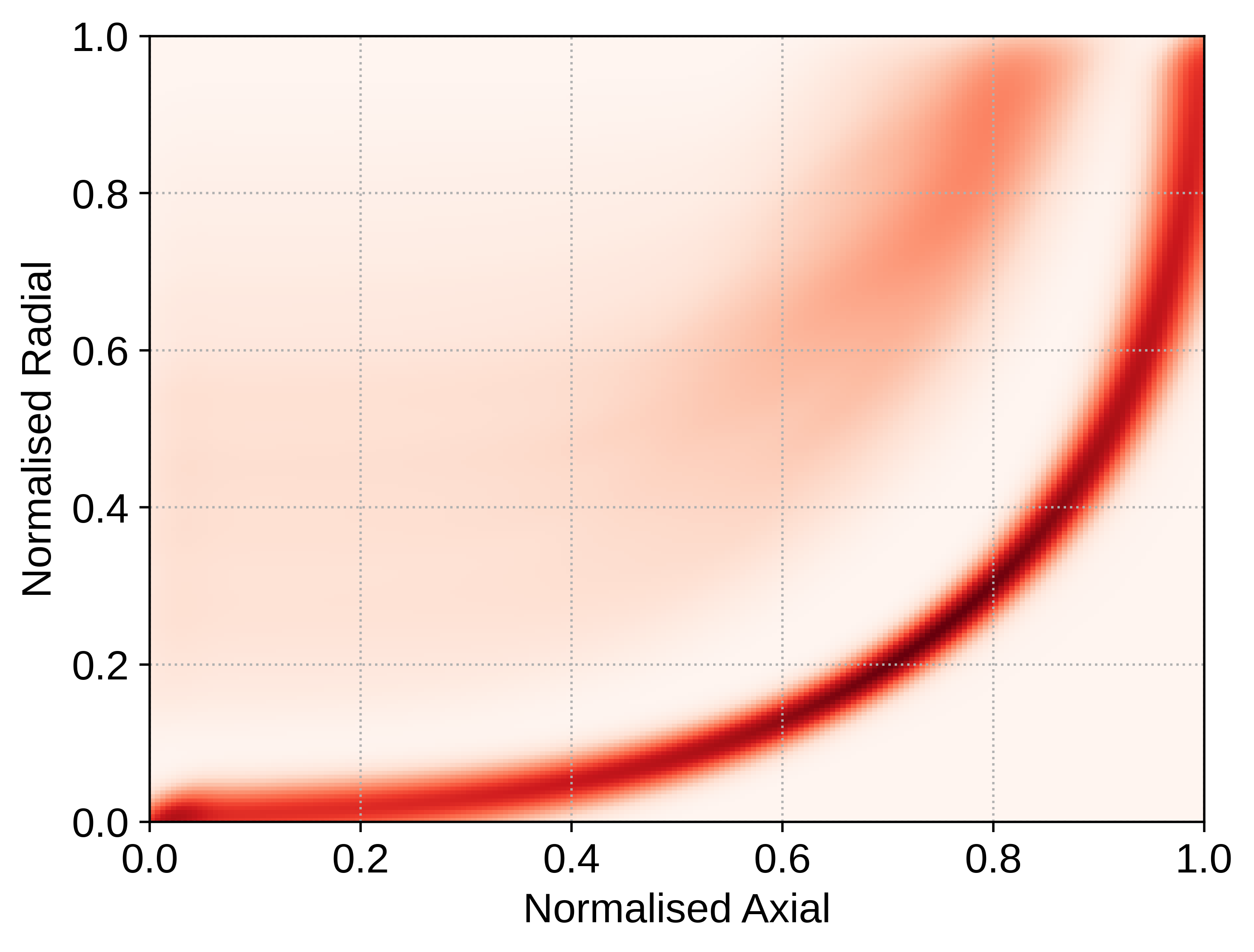}
    \caption{Kernel density estimate of the normalised compressor geometry distribution (normalised meridional view $\overline{r}$ vs. $\overline{x}$)}
    \label{fig:meridional_view_distribution}
\end{figure}

The second consideration for 3D blade coordinate normalisation is the suitability of direct global min-max normalisation. Due to the variation of compressor sizes, if the global normalisation using Eq.~\eqref{eq:normalisation} is applied directly, the normalised coordinates of big compressors will be close to unity, whereas those values for small compressors will be close to zero. In addition, since the compressor performance tends to have a strong correlation with its size, the data weighting issue still exists, making the forthcoming training process more difficult. 

Therefore, instead of using global normalisation in Cartesian coordinates, a two-step normalisation approach in cylindrical polar coordinates is adopted in this study. Each of the compressor blades is firstly normalised to $[0-1]$ using its own min-max values of the three axes ($x,r,\theta$), as shown by Eqs.~\eqref{equation: x norm}-\eqref{equation: theta norm}, where the subscripts min and max are different for each blade design, not a global value.
\begin{equation}
    \label{equation: x norm}
    \overline{x} = \dfrac{x - x_{\text{min}}}{x_{\text{max}}-x_{\text{min}}}
\end{equation}
\begin{equation}
    \label{equation: r norm}
    \overline{r} = \dfrac{r - r_{\text{min}}}{r_{\text{max}}-r_{\text{min}}}
\end{equation}
\begin{equation}
    \label{equation: theta norm}
    \overline{\theta} = \dfrac{\theta - \theta_{\text{min}}}{\theta_{\text{max}}-\theta_{\text{min}}}
\end{equation}
The blade geometry in the normalised $\overline{x},\overline{r},\overline{\theta}$ coordinate system is illustrated by Fig.~\ref{subfig:cylindrical_polar_coordinate_normalsied_blade}. By doing so, each blade is firstly locally scaled such that all normalised coordinates lie within a unified three-dimensional unit bounding box, thereby eliminating the magnitude difference between samples. To make them visually distinguishable, the blade in the normalised $\overline{x},\overline{r},\overline{\theta}$ coordinate is shown with discretised points in blue in this paper.

The final normalised geometries are then plotted on the meridional view and shown as a KDE plot in Fig.~\ref{fig:meridional_view_distribution}. Due to the definition of the cylindrical polar coordinate specification, the inlet and outlet hub points are always (0,0) and (1,1) on the meridional view. Thus, the KDE distribution of the hub profile is much denser in Fig.~\ref{fig:meridional_view_distribution}.

The bounding-box min/max information, or the local scaling parameters namely $x_{\text{min}}$, $x_{\text{max}}$, $r_{\text{min}}$, $r_{\text{max}}$, $\theta_{\text{min}}$ and $\theta_{\text{max}}$ from Eqs.~\eqref{equation: x norm}-\eqref{equation: theta norm}, are also collected for each compressor during the local normalisation, and are treated as the additional geometry parameter. These parameters were then normalised using their corresponding global min/max values as Eq.~\eqref{eq:normalisation} for the following training process.

Up to this point, all the training dataset generation is complete for both geometry and performance data. The dataset is split into training, validation, and testing with a ratio of 8:1:1.

\subsection{Model Implementation}
\label{subsection:model_implementation}
\subsubsection{Elucidated Diffusion Model}
\label{subsubsection:elucidating_diffusion_model}

A conditional diffusion model serves as the generative backbone of this work, following the work in~\cite{Geng2026}. At a conceptual level, the model functions as a denoising estimator that takes as input a noisy geometry representation, a conditioning vector containing the operating or design parameters, and the corresponding noise level, and predicts the underlying clean geometry signal. This denoising process is realised by a neural network, denoted by $\mathcal{D}$, whose architecture is described in Section~\ref{subsubsection:unet_model_architecture}.

The diffusion formulation adopted in this study is based on the Elucidated Diffusion Model (EDM) framework proposed by \citet{Karras2022}. EDM provides a unified reformulation of both the Denoising Diffusion Probabilistic Model (DDPM)~\cite{Ho2020} and score-based generative models~\cite{Song2021}, leading to improved sampling efficiency and overall generation performance. Within this framework, the diffusion process is characterised by the probability flow ordinary differential equation (ODE) and the stochastic differential equation (SDE), as expressed in Eq.~\eqref{edm equation}~\cite{Song2021}.
\begin{equation}
    \label{edm equation}
    \mathrm{d}\mathbf{x}_{\pm} = \underbrace{-\dot{\sigma}(t) \sigma(t) \nabla_{\mathbf{x}} \log p_t (\mathbf{x}; \sigma(t)) \mathrm{d}t}_{\text{probability flow ODE}} 
    \pm 
    \underbrace{\beta(t) \sigma(t)^2 \nabla_{\mathbf{x}} \log p_t (\mathbf{x}; \sigma(t)) \mathrm{d}t + \sqrt{2\beta(t)} \sigma(t) \mathrm{d}\omega_t}_{\text{Langevin diffusion SDE}}
\end{equation}

The first term, corresponding to the deterministic probability flow ODE, drives the diffusion trajectory towards regions of higher data likelihood. The second term in Eq.~\eqref{edm equation} represents the Langevin SDE, which injects stochasticity into the sampling process to enhance the diversity of the generated outputs. In practice, this stochastic behaviour is realised through the temporary addition of Gaussian noise during sampling.

In Eq.~\eqref{edm equation}, $t \in [0, T]$ is the continuous diffusion time variable, $\mathbf{x}$ denotes the data, $p_t$ represents the data probability distribution, and $\sigma$ is the signal standard deviation that characterises the noise level. The score function, $\nabla_x \log p_t(x; \sigma(t))$, is defined in Eq.~\eqref{diffusion score}~\cite{Song2021} and corresponds to the gradient of the log-probability density, directing the diffusion process towards regions of higher data density for the specified noise level ($\sigma(t)$).
\begin{equation}
    \label{diffusion score}
    \nabla_{\mathbf{x}} \log p_t (\mathbf{x}; \sigma(t)) = \frac{D(\mathbf{x};\sigma)-\mathbf{x}}{\sigma^2}
\end{equation}

Following the Elucidated Diffusion Model (EDM) framework~\cite{Karras2022}, the network is trained by minimizing the expected weighted mean squared error between the predicted output and the corresponding ground-truth signal. The weighting function, \(\lambda(\sigma)\), together with the scaling coefficients \(c_{\text{in}}(\sigma)\), \(c_{\text{out}}(\sigma)\), \(c_{\text{skip}}(\sigma)\), and \(c_{\text{noise}}(\sigma)\), are noise-dependent functions specified by the EDM formulation. The resulting training objective is to minimize the expected loss defined in Eq.~\eqref{eq:loss_function}.
\begin{equation}
\label{eq:loss_function}
\mathbb{E}_{\sigma, x, n} \Big[
\underbrace{\lambda(\sigma)c_{\text{out}}^2}_{\text{weight}}
\Big\lVert
\underbrace{F_\theta\!\left(c_{\text{in}}(\sigma)(x+n),\, c_{\text{noise}}(\sigma)\right)}_{\text{neural network output}}
-
\underbrace{\frac{1}{c_{\text{out}}(\sigma)}
\left(x - c_{\text{skip}}(\sigma)(x+n)\right)}_{\text{ground truth}}
\Big\rVert_2^2
\Big]
\end{equation}

After training, the sampling procedure follows the denoising algorithm outlined in Algo.~\ref{edm generation code}. The process begins by initializing the signal, \(\mathbf{x}_0\), as pure Gaussian noise. The diffusion model then progressively removes noise from the signal until the final clean sample, \(\mathbf{x}\), is obtained. The noise level at each denoising step is determined by the schedule$\left( \sigma_{\text{max}}^{\frac{1}{\rho}} + \frac{i}{N-1}\left(\sigma_{\text{min}}^{\frac{1}{\rho}}-\sigma_{\text{max}}^{\frac{1}{\rho}}\right)\right)^{\rho}$, where \(N\) is the total number of denoising steps, \(i\) is the current sampling step, and \(\rho\) is a tunable hyper-parameter. As part of the sampling procedure, a temporary Gaussian perturbation is introduced (line 6 in Algo.~\ref{edm generation code}) to inject additional stochasticity and thereby increase the diversity of the generated designs. Furthermore, unlike the original DDPM sampling algorithm~\cite{Ho2020}, the EDM framework employs a second-order numerical solver (lines 9--12 in Algo.~\ref{edm generation code}) to integrate the denoising trajectory. This improves sampling efficiency by reducing the number of denoising steps required to generate high-quality designs. All hyper-parameters are set to the default values recommended by \citet{Karras2022}.

\begin{algorithm*}
    \caption{Diffusion model generation}
    \label{edm generation code}
    \begin{algorithmic}[1]
    \State \textbf{procedure} Stochastic Sampler ($D (\mathbf{x}; \sigma), t_{i\in \{0,\, ...,\, N \}}, \gamma_{i\in \{0,\, ...,\, N-1 \}}, S_{\text{noise}}$) 
    \State \textbf{sample} $\mathbf{x}_0 \sim \mathcal{N}(\boldsymbol{0}, t_0^2\,\mathbf{I})$
    \For {$i\in \{ 0,\, ...,\, N-1\}$}
        \State \textbf{sample} $\boldsymbol{\epsilon}_i \sim \mathcal{N}(\boldsymbol{0}, S_{\text{noise}}^2\,\mathbf{I})$
    
        \State $\gamma_i = \begin{cases}
            \text{min} \left( \frac{S_{\text{churn}}}{N}, \sqrt{2}-1 \right) &\text{if}~t_i \in [S_{\text{tmin}}, S_{\text{tmax}}] \\
            0 &\text{otherwise}
        \end{cases}$
        
        \State $\hat{t}_i \gets t_i + \gamma_it_i$ \Comment{add noise temporarily to emulate the stochastic diffusion}
        \State $\hat{\mathbf{x}}_i \gets \mathbf{x}_i + \sqrt{\hat{t}_i^2 - t_i^2} \, \boldsymbol{\epsilon}_i$
        \State $\mathbf{d}_i \gets (\hat{\mathbf{x}}_i - D (\hat{\mathbf{x}}_i; \hat{t}_i))/\hat{t}_i$
        \State $\mathbf{x}_{i+1} \gets \hat{\mathbf{x}}_i + (t_{i+1} - \hat{t}_i) \, \mathbf{d}_i$ \Comment{linear Euler step for solving probabilistic flow ODE}
    
        \If {$t_{i+1} \neq 0$}
            \State $\mathbf{d}_i' \gets (\mathbf{x}_{i+1} - D (\mathbf{x}_{i+1}; t_{i+1})) / t_{i+1}$ \Comment{Huen method correction}
            \State $\mathbf{x}_{i+1} \gets \hat{\mathbf{x}}_i + \frac{1}{2}(t_{i+1} - \hat{t}_i)(\mathbf{d}_i + \mathbf{d}_i')$
        \EndIf
    \EndFor
    \end{algorithmic}
\end{algorithm*}

% $\mathbf{x} \subset \mathbb{R}^{3 \times 16 \times 512}$
\subsubsection{Serial Models}
\label{subsubsection:serial_models}
As mentioned in Section~\ref{subsubsection:data_normalisation}, the 3D compressor blade coordinates are individually normalised, and hence the bounding box min/max information is required to transform the model-generated geometry in $(\overline{x}, \overline{r}, \overline{\theta})$ to dimensional cylindrical polar coordinates in $(x,r,\theta)$. In addition, the number of blades is needed to construct a full centrifugal compressor. In this study, we propose a flexible, dual diffusion model architecture, as illustrated by the model training and deployment blocks in Fig.~\ref{fig:work_flow_chart}. This framework employs two diffusion models, namely the main model $\mathcal{M}_{\text{main}}$ and the auxiliary model $\mathcal{M}_{\text{aux}}$, to generate centrifugal compressor geometries. The main model $\mathcal{M}_{\text{main}}$ is trained to produce full 3D blade coordinates $\boldsymbol{x}_{\text{main}} \in \mathbb{R}^{3 \times 16 \times 512}$, which is 3 channels in the normalised coordinates $(\overline{x}, \overline{r}, \overline{\theta})$, 16 span-wise profiles and 512 discretisation points. The network is conditioned on a vector $\boldsymbol{c}_{\text{main}}$ defined as Eq.~\eqref{equation:condition_main}, 
\begin{equation}
    \label{equation:condition_main}
    \boldsymbol{c}_{\text{main}} = \left[ \overline{\dot{m}}, \overline{\omega}, \overline{\eta}, \overline{\textrm{PR}}, \textrm{Flatten} \left( \boldsymbol{g} \right) \right]^{\mathsf{T}} \in \mathbb{R}^{11}
\end{equation}
comprising the normalised desired operating conditions ($\dot{m}$, $\omega$), the desired performance ($\eta$, $\textrm{PR}$), and the flattened compressor geometry descriptor vector $\boldsymbol{g}$, which includes the bounding box min/max values and the number of blades as shown by Eq.~\eqref{equation:g_main}.
\begin{equation}
    \label{equation:g_main}
    \boldsymbol{g} = \left [ \overline{x_{\text{min}}}, \overline{x_{\text{max}}}, \overline{r_{\text{min}}}, \overline{r_{\text{max}}}, \overline{\theta_{\text{min}}}, \overline{\theta_{\text{max}}}, \overline{Z} \right]^{\mathsf{T}}
\end{equation}

The geometry descriptor vector $\boldsymbol{g}$ can either be manually specified or obtained from the auxiliary model $\mathcal{M}_{\text{aux}}$, producing two deployment modes, the automatic and the manual modes, as shown in Fig.~\ref{fig:work_flow_chart}. 
In the fully automatic compressor design generation framework, an additional model of $\mathcal{M}_{\text{aux}}$ is trained and used to produce the geometry descriptor vector $\boldsymbol{g}$. It uses the same EDM framework as introduced in Section~\ref{subsubsection:elucidating_diffusion_model}, but now with a multi-layer perceptron (MLP) architecture and $\boldsymbol{x}_{\text{aux}} = \boldsymbol{g} \in \mathbb{R}^{7}$. The condition of the auxiliary model is as Eq.~\eqref{equation:condition_aux}
\begin{equation}
    \label{equation:condition_aux}
    \boldsymbol{c}_{\text{aux}} = \left[ \overline{\dot{m}}, \overline{\omega}, \overline{\eta}, \overline{\textrm{PR}} \right]^{\mathsf{T}} \in \mathbb{R}^{4}
\end{equation}
The generated $\boldsymbol{g}$ is then fed to the main model as part of the condition $\boldsymbol{c}_{\text{main}}$ as shown by Eq.~\eqref{equation:condition_main}. The generated non-dimensional geometry $\boldsymbol{x}_{\textrm{main}}$ can hence be retrieved and denormalised into $\boldsymbol{X}_{\textrm{blade}}$, which maps $(\overline{x}, \overline{r}, \overline{\theta})$ back to dimensional coordinates $(x,r,\theta)$ using the bounding box min/max values stored in $\boldsymbol{g}$.
For certain applications with a geometry constraint (e.g., maximum radius), the auxiliary model can be detached from the main model, and the geometry descriptor vector can be manually specified to provide operational flexibility. 

% The model architecture is much simpler for the 1D-based diffusion model. Since the 1D parametrised geometries are globally normalised, there is no additional geometry descriptor vector needed, and the standard single-diffusion model architecture is sufficient. 

In summary, the overall blade design generation in Fig.~\ref{fig:work_flow_chart} proposed in this paper can be written as Eq.~\eqref{equation:dual_model_workflow}, providing a simple and flexible workflow.
\begin{equation}
    \label{equation:dual_model_workflow}
    \begin{split}
        \boldsymbol{X}_{\text{blade}} = 
        \mathrm{Denorm} \Bigg(
        \mathcal{M}_{\text{main}}\Big(
        \big[\overline{\dot{m}}, \overline{\omega}, \overline{\eta}, \overline{\textrm{PR}}, \mathrm{Flatten}(\boldsymbol{g})\big]^{\mathsf{T}}
        \Big)
        \Bigg), \\
        \quad
        \boldsymbol{g} = 
        \begin{cases} 
        \mathcal{M}_{\text{aux}}\big([\overline{\dot{m}}, \overline{\omega}, \overline{\eta}, \overline{\textrm{PR}}]^{\mathsf{T}}\big) \;\; \text{(automatic mode)},\\[1mm]
        \text{Specified (manual mode)}.
        \end{cases}
    \end{split}
\end{equation}

\begin{figure*}
    \centering
    \includegraphics[width=\textwidth]{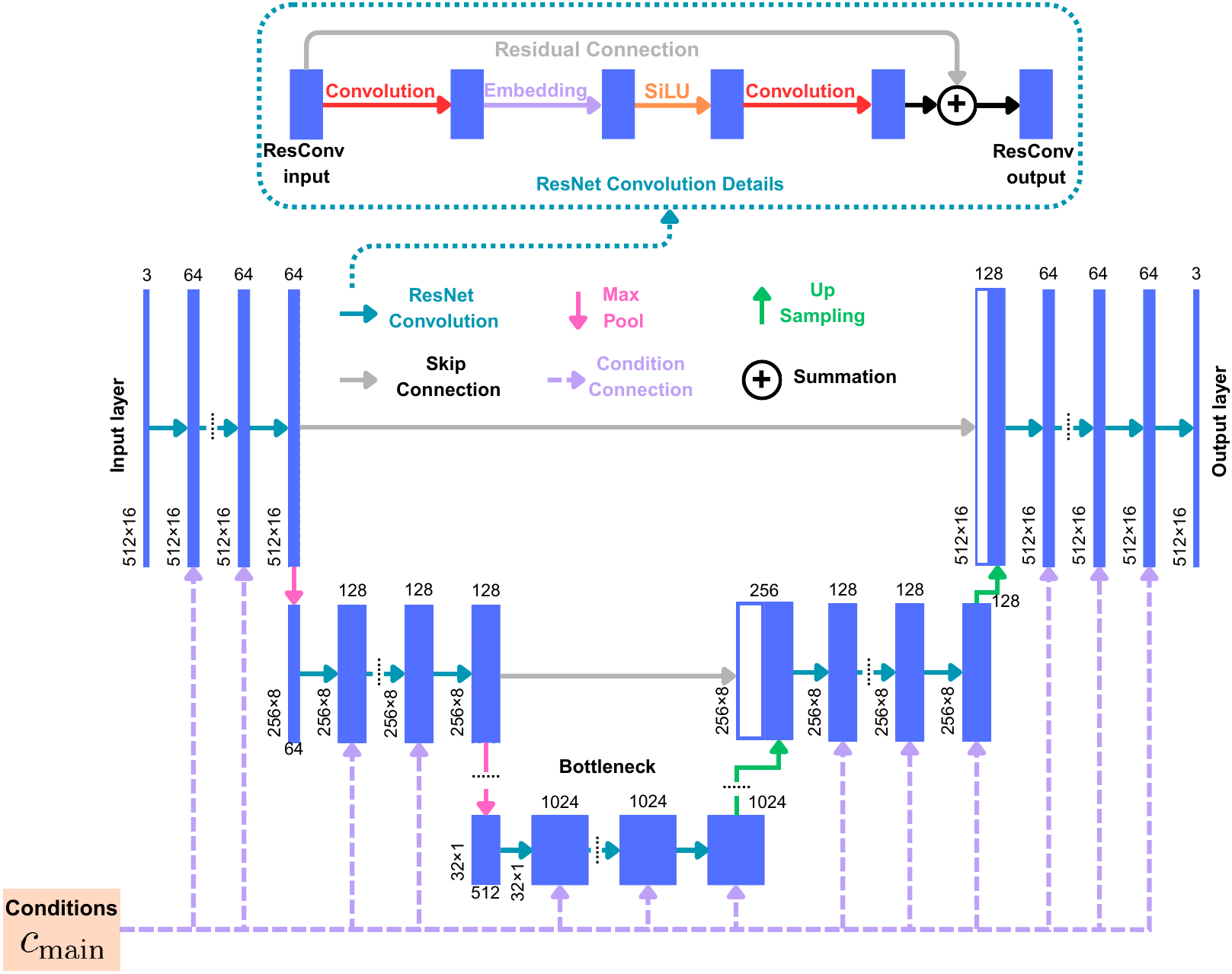}
    \caption{UNet Architecture for Denoising Neural Network}
    \label{fig:unet_architecture}
\end{figure*}

In addition, to better compare the diffusion model performance with different geometry encoding methods, a reference 1D diffusion model $\mathcal{M}_{\text{1D}}$ is trained. The data for this model is the 1D geometry parametrised geometries as shown in Table~\ref{table:1D_geometry_parameters} and conditioned with operating and performance parameters ($\boldsymbol{c}_{\text{1D}}=\boldsymbol{c}_{\text{aux}}$ from Eq.~\eqref{equation:condition_aux}). This model presents a typical framework in the literature that is trained with parametrised geometries, hence reproduced in this work for comparison.

\subsubsection{Denoiser Network Architecture}
\label{subsubsection:unet_model_architecture}

As introduced in Section~\ref{subsubsection:elucidating_diffusion_model}, a denoiser neural network $D\brac{\boldsymbol{x}, \boldsymbol{c}, t}$ is needed for each diffusion model $\mathcal{M}\brac{\boldsymbol{c}}$ as the core of the sampling process shown in Algo.~\ref{edm generation code}. For this paper with the EDM framework, the denoiser network can be presented as Eq.~\eqref{equation:general_denoiser}, where $\mathcal{X}$ is the input and output space, $\mathcal{C}$ is the conditioning space, and $\{0, \dots, T\}$ is the discrete diffusion time steps.
\begin{equation}
    \label{equation:general_denoiser}
    D : \mathcal{X} \times \mathcal{C} \times \{0, \dots, T\} \rightarrow \mathcal{X}
\end{equation}

\paragraph{Main Model}

In this paper, U-Net~\cite{Ronneberger2015} is implemented as the step-level denoiser network for the main model $\mathcal{M}_{\text{main}}$. Since the geometry tensor $\boldsymbol{x}_{\text{main}} \in \mathcal{X}_{\text{main}} \subset \mathbb{R}^{3 \times 16 \times 512}$ can be viewed as a 3-channel image, enabling the use of a U-Net, which is the most commonly adopted network architecture in diffusion models~\cite{Ho2020, Liu2023}. In addition, residual connections (inspired by ResNet~\cite {He2015}) are adopted as the basic convolutional building blocks. The overall architecture of $D_{\text{main}}$ is illustrated in Fig.~\ref{fig:unet_architecture}.

Firstly, the input data $\boldsymbol{x}_{\text{main}_i}$ are mapped to a 64-channel feature map via a lifting convolution layer. Then the data are progressively downsampled via the encoder, which consists of a series of residual convolution blocks and max-pooling layers. For each ResNet block, the spatial resolution and number of feature channels remain unchanged, while each max-pooling layer halves the spatial resolution and doubles the feature channels. 

After the bottleneck, as a symmetric architecture, the data are upsampled to gradually restore their original resolution while reducing the feature channels. For each upsampling layer, the data resolution is doubled while the feature channel is halved. The upsampled features are then concatenated with the corresponding features from the encoder via skip connections, enhancing the fusion of low- and high-resolution information and helping to recover details lost during downsampling.

The condition vector $\boldsymbol{c}_{\text{main}} \in \mathcal{C}_{\text{main}} \subset \mathbb{R}^{11}$ and the sampling time step $t$ are firstly encoded using sinusoidal positional embedding, followed by linear projection and nonlinearity. The resulting embeddings are combined via element-wise addition to form a unified conditioning embedding before being injected into each ResNet block in the U-Net as illustrated in Fig.~\ref{fig:unet_architecture}. This ensures that the model is consistently constrained by the conditions. 

% Although the models trained in this study share the same basic U-net architecture, the detailed neural network architecture is different. The U-net architecture details for different diffusion models are specified in Table~\ref{table:neuro_network_architecture_details}

\paragraph{Auxiliary Model}

For the auxiliary model $\mathcal{M}_{\text{aux}}$, the same EDM framework is adopted, but with a different network architecture. Specifically, the denoiser is implemented as a multi-layer perceptron (MLP) with a depth of $L=10$ and hidden layer width of 512. The model operates on the geometry descriptor vector $\boldsymbol{g} \in \mathcal{X}_{\text{aux}} \subset \mathbb{R}^{7}$, rather than spatially structured data.

The conditioning vector of the auxiliary model $\boldsymbol{c}_{\text{aux}} \in \mathcal{C}_{\text{aux}} \subset \mathbb{R}^{4}$ and the diffusion time step $t$ are encoded in the same manner as in the main model, using sinusoidal positional embeddings followed by linear projection and nonlinearity. The resulting embeddings are combined via element-wise addition to form a unified conditioning embedding, albeit with a different embedding dimension tailored to the MLP architecture.

\paragraph{1D Model}

Similarly, the reference 1D model $\mathcal{M}_{\text{1D}}$ uses the EDM framework and implements the denoiser as an MLP with a depth of $L=10$ and hidden layer width of 512. The model operates on the 1D geometry parameter vector $\boldsymbol{x}_{\text{1D}} \in \mathcal{X}_{\text{1D}} \subset \mathbb{R}^{12}$ as a lower-order representation of the compressor geometry (parameters in Table~\ref{table:1D_geometry_parameters}).

The conditioning vector of the 1D model is the same as the auxiliary model $\boldsymbol{c}_{\text{1D}} \in \mathcal{C}_{\text{1D}} \subset \mathbb{R}^{4}$ and the diffusion time step $t$ are encoded in the same manner as in and $\mathcal{M}_{\text{aux}}$.

\begin{table*}[htbp]
    \centering
    \caption{The neural network architecture details for different diffusion models}
    \label{table:neuro_network_architecture_details}
    \begin{tabular}{lccc}
        \hline
        \textbf{} & \textbf{1D Parametrised Geometry Model} & \textbf{3D Auxiliary Model} & \textbf{3D Main Model} \\
        \hline
        Building block & MLP & MLP & Conv2D\\
        Number of downsampling layers & 1 & 1 & 5 \\
        Number of ResNet block per layer & 10 & 10 & 5 \\
        Model parameter numbers & 52,470,286 &52,463,111 &166,597,123  \\
        Training batch size & 64 & 64 & 32\\
        Training epochs & 100 & 100 & 300 \\
        Training time & 17 min & 16 min & 36 h\\
        \hline
    \end{tabular}
\end{table*}

\subsection{Model Deployment}
\label{subsection:model_deployment}

\subsubsection{Single-Target Generation}
\label{subsubsection:single_target_generation}

The trained diffusion model can then be used to generate compressor geometries using Algo.~\ref{edm generation code} with target operating conditions ($\dot{m}_{\text{target}}$ and $\omega_{\text{target}}$) and the performance targets ($\eta_{\text{target}}$ and $PR_{\text{target}}$). The denormalised blade geometry $\boldsymbol{X}_{\textrm{blade}}$ can then be retrieved using the geometry descriptor vector $\boldsymbol{g}$, which is either directly specified or generated using $\mathcal{M}_{\text{aux}}$ as shown in Eq.~\eqref{equation:dual_model_workflow}. The ground truth of the performance values, $\eta_{\text{actual}}$ and $PR_{\text{actual}}$, of the generated geometries are evaluated using the meanline model here again, for faster performance evaluation and for data consistency. Measurements are taken to extract the 1D geometry parameters from the 3D geometry, as the meanline model input.  The targeted and actual performances are then compared for design accuracy, which is evaluated as an average percentage deviation from the ground truth values as Eq.~\eqref{equation:error}.
\begin{equation}
    \label{equation:error}
    \begin{split}
        \textrm{RE}_{\textrm{PR}} &= \left|\frac{\textrm{PR}_{\text{actual}} - \textrm{PR}_{\text{target}}}{\textrm{PR}_{\text{target}}}\right| \times 100 \,\% \\
        \textrm{RE}_{\eta} &= \left|\frac{\eta_{\text{actual}} - \eta_{\text{target}}}{\eta_{\text{target}}}\right| \times 100 \,\%
    \end{split}
\end{equation}

% However, the performance of the compressor is highly sensitive to its geometry, where small geometric differences can result in significant variations in aerodynamic characteristics. In addition, the stochastic nature of the diffusion sampling process may introduce geometry artefacts. 

The diffusion model is trained to approximate the conditional distribution of compressor geometries associated with a prescribed aerodynamic target. Under the manifold hypothesis, physically feasible compressor geometries occupy a low-dimensional subset of the high-dimensional representation space of geometries. Therefore, for a given design target, the corresponding solutions are assumed to form a conditional distribution supported on a subset of the compressor geometry manifold. However, due to finite training data and model approximation errors, the learned distribution generally differs from the true conditional distribution. Furthermore, given that the aerodynamic performance of a compressor tends to be very sensitive to its geometry, the model-generated geometries may not always achieve the desired $\textrm{PR}$ and $\eta$ values. Consequently, the generated geometries do not always achieve the desired $\eta$ and $\textrm{PR}$ values, and multiple sampling trials may be required to satisfy the target values within a specified tolerance $\varepsilon$. 

\begin{figure*}
    \centering
    \begin{subfigure}{0.49\textwidth}
    \centering
    \includegraphics[width=0.6\textwidth]{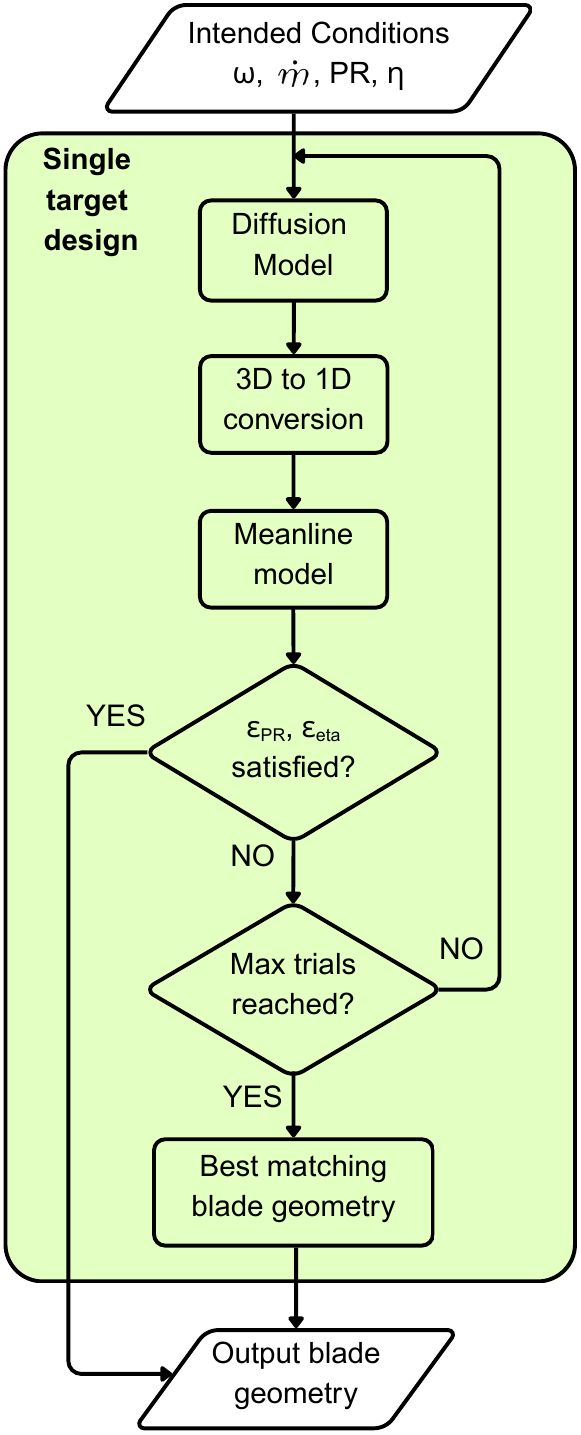}
    \caption{Single-Target Generation}
    \label{subfig:single_target_generation_flow chart}
    \end{subfigure}
    \begin{subfigure}{0.49\textwidth}
    \centering
    \includegraphics[width=0.711\textwidth]{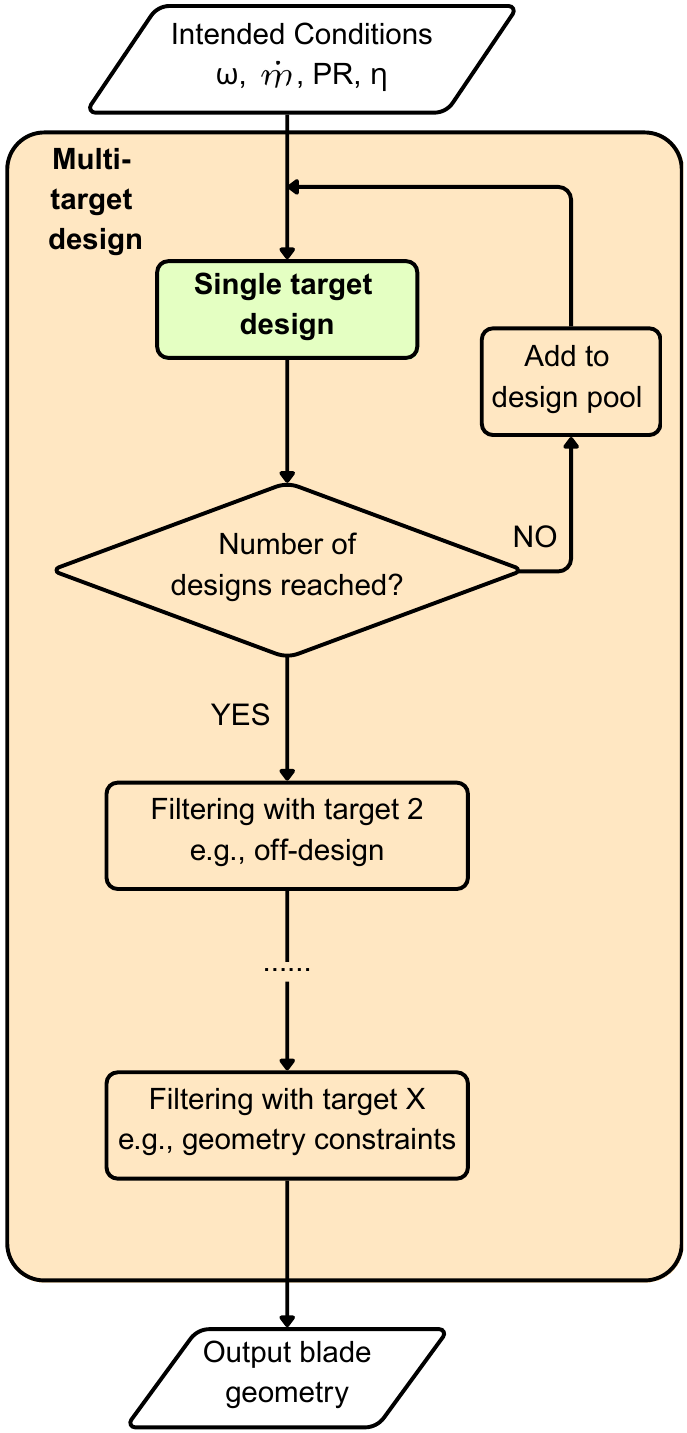}
    \caption{Multi-Target Generation}
    \label{subfig:multi_target_generation_flow_chart}
    \end{subfigure}
    \caption{Proposed Design Generation Procedures of Diffusion Model}
    \label{fig:generation_flow_charts}
\end{figure*}

Therefore, the single-target geometry generation procedure used in this paper is illustrated by Fig.~\ref{fig:generation_flow_charts}, following a similar procedure proposed in~\cite{Geng2026}. A tolerance $\varepsilon$ is predefined for each performance target ($\eta$ and $\textrm{PR}$) and is used as the acceptance criterion to determine whether a generated blade geometry is considered a valid design (when $\textrm{RE} \leq \varepsilon$). If no generated designs can satisfy $\textrm{RE} \leq \varepsilon$ after the max number of trials $N_{\text{trial}}$, the design with the least \textrm{RE} will be selected and returned. 

As the tolerance $\varepsilon$ becomes tighter, the inverse design problem becomes increasingly challenging, requiring a greater number of sampling trials to obtain a satisfactory design candidate. For demonstrations presented in this paper, the tolerances for both $\eta$ and $\textrm{PR}$ are set to be 1\,\%, and a maximum of 100 trials is allowed for model evaluations. 

\subsubsection{Multiple-Target Generation}
\label{subsubsection:multiple_target_generation}

As introduced in Section~\ref{section:introduction}, the stochastic nature of diffusion models is particularly advantageous for inverse design. It enables the effective exploration of the design space and provides diverse candidate design solutions, which align naturally with the objectives of the inverse design problems, where multiple distinct geometries can satisfy the same performance targets.

This property also makes the diffusion model suitable for multi-objective design tasks. The multi-target design generation procedure is presented in Fig.~\ref{subfig:multi_target_generation_flow_chart} following a similar procedure in~\cite{Geng2026}. During this process, the single-target generation procedure is deployed multiple times to obtain a variety of candidate compressor geometries. This forms a pool of designs from which further filtering can be conducted according to additional specific design requirements, which may include minimum blade thickness for structural concerns and off-design performances. 

\subsection{Model Evaluation Metrics}
\label{subsection:model_evaluation_metrics}

In order to systematically evaluate the design capability of the trained diffusion model, several metrics are introduced in this section describing the model performance from various aspects.

\subsubsection{Design Accuracy}
\label{subsubsection:design_accuracy}

The first and most important requirement of the diffusion model is to generate a compressor geometry that can satisfy the specified design target with high accuracy. In this paper, the design accuracy is quantified by the root-mean-square relative error (RMSE) of the two performances metrics ($\textrm{PR}$ and $\eta$) of the generated design solutions, as shown by Eq.~\eqref{eq:RMSE}, where subscripts target and actual represents the corresponding values that were specified model condition inputs and from the ground truth calculated using the meanline model (Section~\ref{subsubsection:single_target_generation}), respectively.
\begin{align}
    \label{eq:RMSE}
    \textrm{RMSE}_{\textrm{PR}} &= \sqrt{\frac{1}{N_{\text{test}}} \sum_{i=1}^{N_{\text{test}}} \left( \frac{\textrm{PR}_{\text{actual}_{i}} - \textrm{PR}_{\text{target}_{i}}}{\textrm{PR}_{\text{target}_{i}}} \right)^2} \notag \\ 
    \textrm{RMSE}_{\eta} &= \sqrt{\frac{1}{N_{\text{test}}} \sum_{i=1}^{N_{\text{test}}} \left( \frac{\eta_{\text{actual}_{i}} - \eta_{\text{target}_{i}}}{\eta_{\text{target}_{i}}} \right)^2}
\end{align}
It is to note that the metrics presented by Eq.~\eqref{eq:RMSE} provide an overall indication of model performance over the whole testing dataset with $N_{\text{test}}$ samples and is different from the single design error shown in Eq.~\eqref{equation:error}.

Two additional parameters are used for model comparison when using the single-target generation procedure illustrated in Fig.~\eqref{subfig:single_target_generation_flow chart}. Firstly, as explained in Section~\ref{subsubsection:single_target_generation}, many generation trials are allowed until either the specified tolerance $\varepsilon$ is satisfied or the maximum number of allowable trials is reached. For a better-performing model, the generated design should achieve a lower RMSE with fewer design trials. Therefore, the average number of design trials is recorded for the testing dataset to quantify the generation efficiency ($N{\text{trial, max}}$ is used if no designs were generated within $\varepsilon$). In addition, the design failure rate, which is defined as the proportion of the physically unfeasible designs (negative volume, meanline model divergence, etc.) is also recorded as part of the design accuracy analysis. 

\subsubsection{Design Diversity}
\label{subsubsection:desisgn_diversity}

As discussed in Section~\ref{subsubsection:multiple_target_generation}, the design diversity is one of the most attractive features of diffusion models for engineering inverse designs. 

From a mathematical perspective, the 3D compressor geometries in the 3D main model lie in a very high-dimensional mathematical space, $\mathbb{R}^{3\times16\times512}$. For a given set of design conditions ($\dot{m}, \omega, \textrm{PR}, \eta$), it is assumed that the subset of geometries satisfying these conditions forms a low-dimensional manifold embedded within this high-dimensional space. This manifold represents the solution set of the inverse design problem, and an effective generative model should be capable of approximating its structure and diversity. 

To assess the extent to which the diffusion model explores this solution manifold, a reference approximation of its structure is required. Directly searching in the $\mathbb{R}^{3\times16\times512}$ space requires a prohibitive amount of computational resources. Therefore, the meanline model in Section~\ref{section: Performance Database Generation} is used to generate a set of solutions for a given design target in a reduced-order space of $\mathbb{R}^{12}$. 

In terms of the actual procedure, the initial design parameters in Table~\ref{table:geometry_sampling_variables} are sampled again using the LHS random sampling, and the remaining 1D compressor parameters are then derived following the same procedure as in Section~\ref{subsubsection:geometry_sampling_and_generation}. These 1D geometry parameters are then fed into the meanline model with a specified operating condition. Designs that satisfy the target performance $(\textrm{PR}, \eta)$ within a prescribed tolerance are filtered out. This process is repeated until 100 feasible designs are obtained. The resulting set of solutions is taken as a discrete, low-dimensional approximation of the underlying solution manifold, which serves as a reference for evaluating the diversity and coverage of the diffusion-based design generation. To ensure this solution set is a good representation of the physical design space and to set a comparison standard, another solution set is generated independently following the same sampling procedures. These two solution sets should have high similarity if both of them are good representations of the full design space for the given condition. These physically sampled solution sets are denoted as sets $\mathcal{B}_1$ and $\mathcal{B}_2$, and they are generally denoted as set $\mathcal{B}$ in this section. 

At the same time, the trained diffusion models are also tasked with the same design target to generate 100 geometry solutions, using the single-target generation procedure (Fig.~\ref{subfig:single_target_generation_flow chart}). These diffusion model generated solution sets are denoted as sets $\mathcal{A}_\textrm{1D}$ and $\mathcal{A}_\textrm{3D}$, and they are generally denoted as set $\mathcal{A}$ in this section. Ideally, the distribution of two solution sets (i.e., sets $\mathcal{A}$ and $\mathcal{B}$) should be matched, indicating that the diffusion model can effectively explore the entire design space. To systematically quantify this, a number of metrics are used and are discussed in the following paragraphs. For clarity, the lowercase $a$ and $b$ represent individual blade geometries in sets $\mathcal{A}$ and $\mathcal{B}$, respectively. 

\textbf{Chamfer distance} Chamfer distance (CD) is a measurement commonly used in computer vision to compare the similarity between images or sets of point clouds~\cite{Barrow1977}. In this study, the blade geometries are represented as structured grids rather than unstructured point clouds. Therefore, the Chamfer distance is defined at the blade level, where the distance between two blades is measured using the root-mean-square (RMS) distance between corresponding grid points, as given in Eq.~\eqref{eq:chamfer_distance}.
\begin{align}
    \label{eq:chamfer_distance}
    \textrm{CD}_{\mathcal{A} \to \mathcal{B}} &= \frac{1}{|\mathcal{A}|} \sum_{a \in \mathcal{A}} \min_{b \in \mathcal{B}} 
    \sqrt{\frac{1}{N} \lVert a - b \rVert_2^2} \notag \\
    \textrm{CD}_{\mathcal{B} \to \mathcal{A}} &= \frac{1}{|\mathcal{B}|} \sum_{b \in \mathcal{B}} \min_{a \in \mathcal{A}} 
    \sqrt{\frac{1}{N} \lVert b - a \rVert_2^2} \notag \\
    \textrm{CD}_{\text{eff}} &= \frac{1}{2} \left( \textrm{CD}_{\mathcal{A} \to \mathcal{B}} + \textrm{CD}_{\mathcal{B} \to \mathcal{A}} \right)
\end{align}
where $N = 512 \times 16$ denotes the number of grid points per blade. For each blade $a \in \mathcal{A}$, the closest blade $b \in \mathcal{B}$ is identified by minimising the RMS distance over all grid points. This process is repeated for all blades in set $\mathcal{A}$, and the average distance is computed. The same procedure is then applied in the reverse direction (subscript $\mathcal{B} \to \mathcal{A}$) to ensure symmetry. The final Chamfer distance is defined as the average of the two directional distances.

Since the metric is based on Euclidean distances in dimensional physical space $(x,y,z)$, the resulting Chamfer distance has units of length.

\textbf{Structural similarity index measure} The structural similarity index measure (SSIM) is another commonly used metric to assess the similarity of structured data such as images~\cite{Wang2004}. In this study, since the blade geometries are represented on structured grids, SSIM can be directly applied to quantify the similarity between two blade geometries.
Similar to the Chamfer distance, SSIM is extended to the set level and evaluated at the blade level, as shown by Eq.~\eqref{eq:ssim}. 
\begin{align}
    \label{eq:ssim}
    \textrm{SSIM}_{\mathcal{A} \to \mathcal{B}} &= \frac{1}{|\mathcal{A}|} \sum_{a \in \mathcal{A}} \max_{b \in \mathcal{B}} \textrm{SSIM}(a,b) \notag \\
    \textrm{SSIM}_{\mathcal{B} \to \mathcal{A}} &= \frac{1}{|\mathcal{B}|} \sum_{b \in \mathcal{B}} \max_{a \in \mathcal{A}} \textrm{SSIM}(a,b) \notag \\
    \textrm{SSIM}_{\text{eff}} &= \frac{1}{2} \left( \textrm{SSIM}_{\mathcal{A} \to \mathcal{B}} + \textrm{SSIM}_{\mathcal{B} \to \mathcal{A}} \right)
\end{align}
For each blade $a \in \mathcal{A}$, the most similar blade $b \in \mathcal{B}$ is identified by maximising the SSIM value, which measures similarity in terms of local structure, contrast, and intensity patterns over the grid representation. This process is repeated for all blades in set $\mathcal{A}$, and the average similarity is computed. The same procedure is then applied in the reverse direction to ensure a symmetric comparison between the two sets.
Although the SSIM metric is symmetric for a given pair of blades, i.e.\ $\mathrm{SSIM}(a,b) = \mathrm{SSIM}(b,a)$, the set-level formulation in Eq.~\eqref{eq:ssim} is inherently asymmetric due to the one-sided matching procedure. Specifically, for each blade in one set, only the most similar blade in the other set is considered. Therefore, $\mathrm{SSIM}_{\mathcal{A} \to \mathcal{B}} \neq \mathrm{SSIM}_{\mathcal{B} \to \mathcal{A}}$ in general. To obtain a symmetric measure between the two sets, the final SSIM is defined as the average of the two directional values.
This formulation assumes a consistent spatial parameterisation of the blade geometries, such that local structural comparisons are meaningful across samples.

The final computed SSIM will be a non-dimensional number in the range $[-1,1]$, with 1 representing identical signals and -1 representing opposite signals. 

\textbf{Jensen-Shannon divergence} Jensen-Shannon divergence (JSD) is a symmetric and smoothed measurement of the similarity between two probability distributions~\cite{Menendez1997}, which is derived from the Kullback–Leibler (KL) divergence. In this study, the 3D KDE is first computed for Sets A and B to construct a continuous approximation of their geometry distribution in the dimensional $(x, y, z)$ coordinate system. The mathematical formulation of KDE follows Eq.~\eqref{eq:kde}.

The JSD used in this study is hence defined as Eq.~\eqref{eq:jsd}, where $p_\mathcal{A}$ and $p_\mathcal{B}$ denote the probability density functions of the geometry distributions corresponding to sets $\mathcal{A}$ and $\mathcal{B}$, respectively.
\begin{align}
    \label{eq:jsd}
    \mathrm{JSD}(p_\mathcal{A}\|p_\mathcal{B}) &=\frac{1}{2}\mathrm{KL}(p_\mathcal{A}\|p_{\text{mix}})+\frac{1}{2}\mathrm{KL}(p_\mathcal{B}\|p_{\text{mix}}) \notag \\
    \text{where  } p_{\text{mix}} &=\frac{1}{2}(p_\mathcal{A}+p_\mathcal{B})
\end{align}

The JSD is a non-negative, symmetric quantity. When the logarithm is taken with base 2, it is bounded in the range $[0,1]$, where 0 indicates identical distributions and 1 indicates maximal dissimilarity.

These three similarity metrics (CD, SSIM, and JSD) are jointly used to quantify and compare the model's design variety. Although all the three metrics can measure the similarity between the two solution sets, CD and SSIM focus more on the solution range coverage, whereas JSD leans more to the solution distribution (i.e., extreme cases appear less frequently in the solution set).

\section{Results}
\label{section:results}
\subsection{Model Design Accuracy Results}
\begin{figure}
    \centering
    \includegraphics[width=\columnwidth]{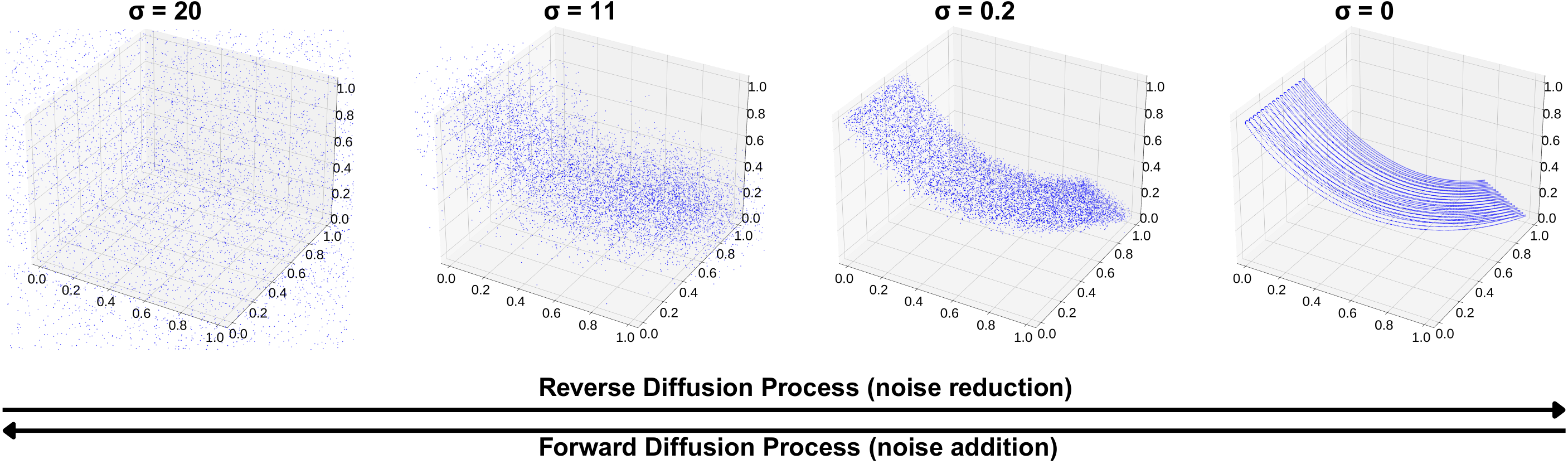}
    \caption{Denoising Process of the 3D Main Diffusion Model}
    \label{fig:denoising_plot}
\end{figure}

\begin{figure*}
    \centering
    \includegraphics[width=\linewidth]{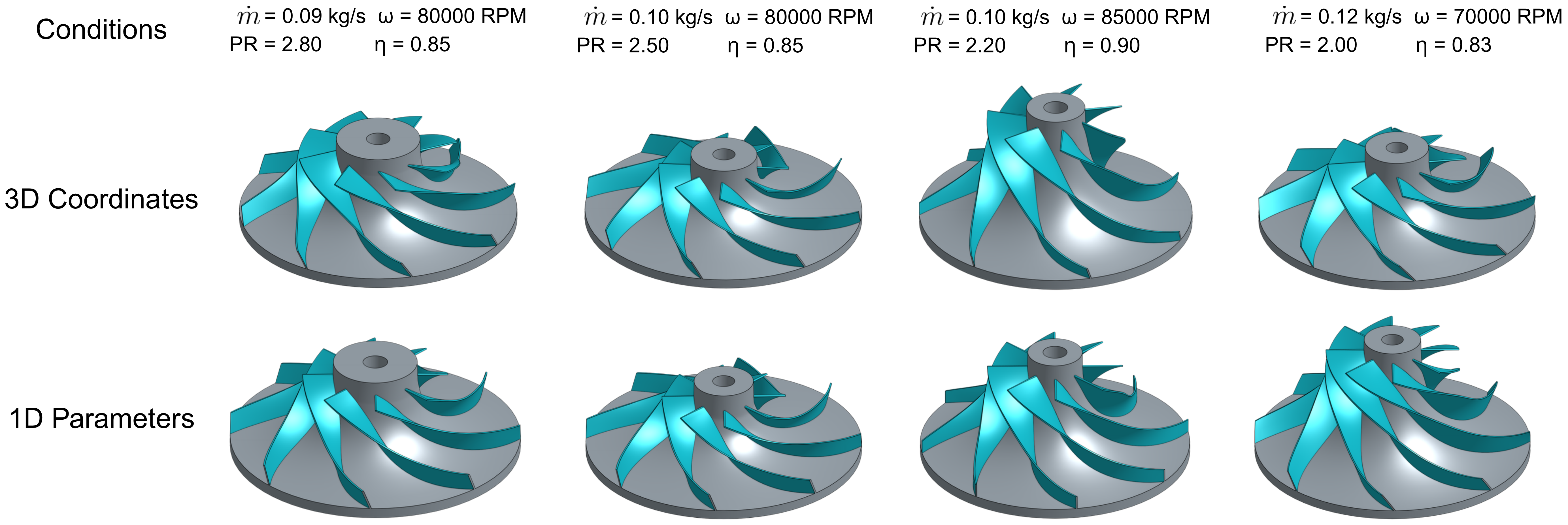}
    \caption{Examples of Generated Designs}
    \label{fig:generation_demo}
\end{figure*}

\begin{figure}
    \centering
    \begin{subfigure}{0.9\textwidth}
    \centering
    \includegraphics[width=\textwidth]{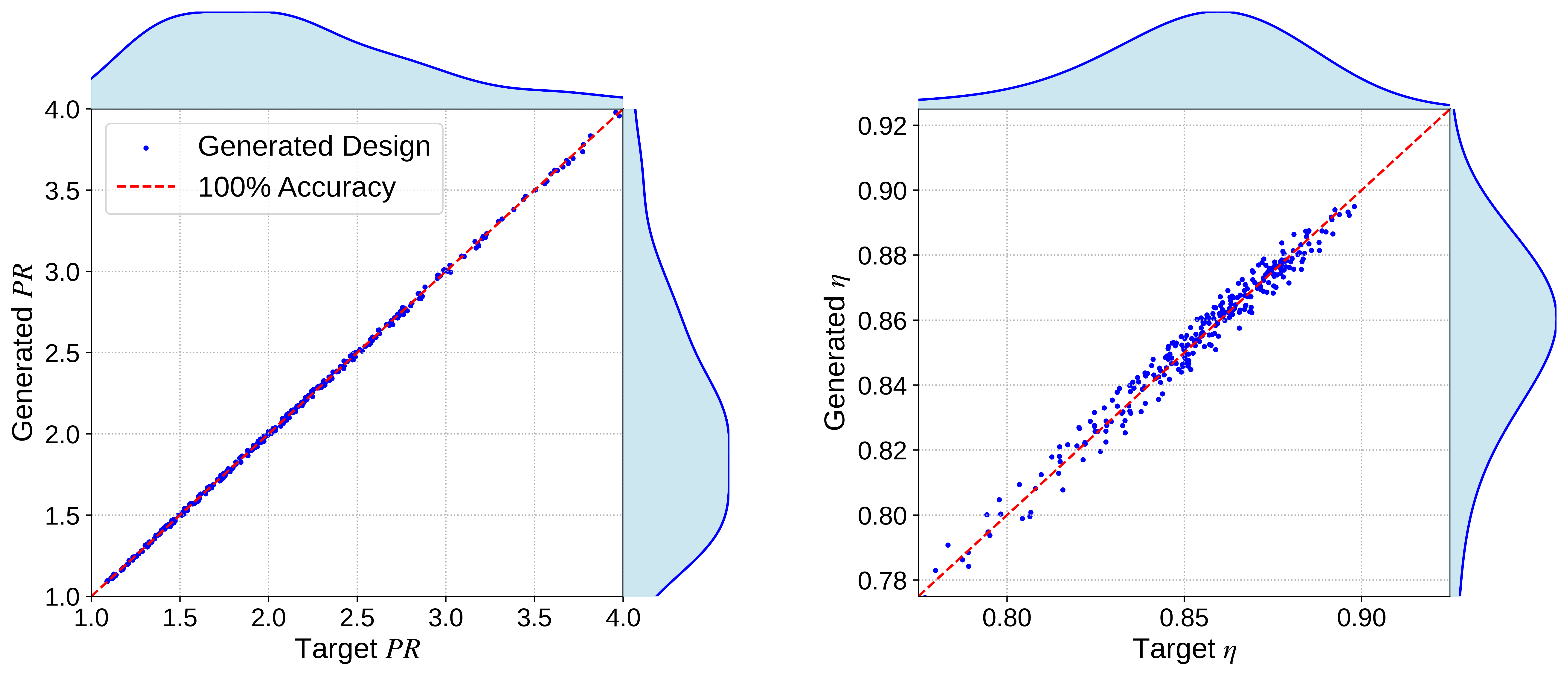}
    \caption{3D Coordinate}
    \label{subfig:accuracy_plot_3D}
    \end{subfigure}
    \begin{subfigure}{0.9\textwidth}
    \centering
    \includegraphics[width=\textwidth]{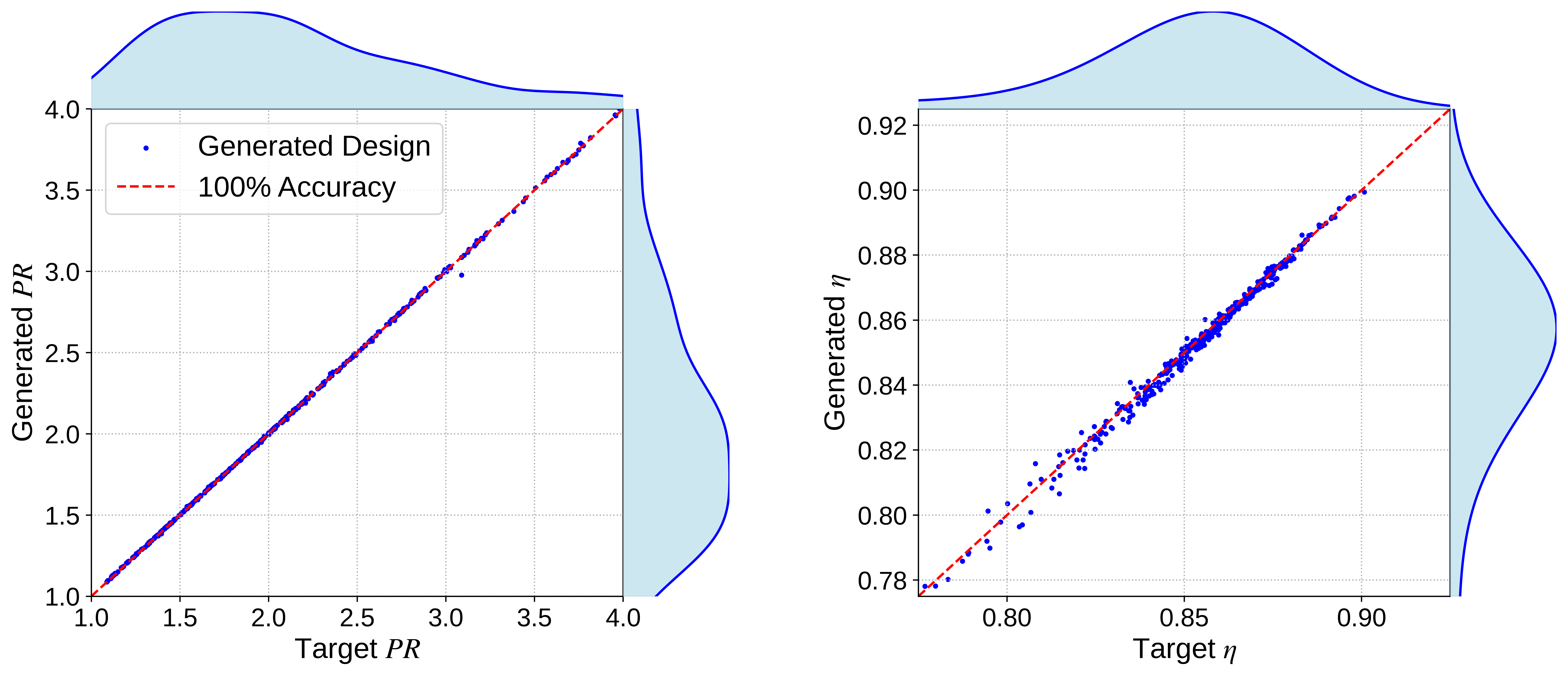}
    \caption{1D Parameterised Geometry}
    \label{subfig:accuracy_plot_1D}
    \end{subfigure}
    \caption{Design Accuracy Plot for Single-Target Generation}
    \label{fig:design_accuracy_plot}
\end{figure}

The sampling process of the 3D main diffusion model is illustrated in Fig.~\ref{fig:denoising_plot}. Four noise levels of $\sigma = 0, 0.2, 11$, and $20$ are displayed to present the forward and reverse diffusion process. % Here the diffusion process only show the 3D main diffusion model in $(\overline{x},\overline{r},\overline{\theta})$, because the data processed by the other two models -- the auxiliary model and 1D baseline diffusion model -- are vectors, which are nothing more than just arrays of numbers and hence cannot be intuitively visualised. 

Four examples of the generated compressor blade profiles from both the 3D diffusion model ($\mathcal{M}_{\text{main}}$ and $\mathcal{M}_{\text{aux}}$) and the 1D baseline diffusion model ($\mathcal{M}_{\text{1D}}$), as shown in Fig.~\ref{fig:generation_demo}. Note that all geometries are denormalised and plotted in Cartesian $(x,y,z)$, and the 1D parametrised geometries generated from the 1D baseline diffusion model are converted to 3D blades using the in-house blade-forming tool for direct visual comparison. These results demonstrate that the 3D diffusion model is able to generate geometrically feasible compressor blades with no significant artefacts. It is noteworthy that the geometry generated directly by the 3D diffusion model exhibits a small amount of pointwise noise, which is a common characteristic in 3D-based diffusion models as reported in other studies~\cite{Li2024, Liu2023}. To improve surface quality, in this study, a polynomial surface fitting procedure is applied, which ensures $\mathcal{C}^1$ surface continuity. The surface fitting process does not significantly alter the geometry shape generated from the diffusion model, with the average distance moved at the order of 0.1\,\% of the compressor axial length ($\mathcal{O}(10 \mu \textrm{m})$). This level of surface roughness is comparable to that observed in typical casting manufacturing processes, and the surface smoothing process can be considered 'virtual polishing'.

\begin{table}
\centering
\caption{Design Accuracy Metrics for Single-Target Generation}
\label{table:design_accuracy_table}
\begin{tabular}{lcc}
\hline
\textbf{Metric} & \textbf{1D Model} &\textbf{3D Model} \\
\hline
$\textrm{PR}$ RMSE & 0.33\,\% & 0.53\,\% \\
$\eta$ RMSE & 0.31\,\% & 0.42\,\% \\
Average number of design trials & 2.02 & 4.45\\
Failure rate & 2.32\,\% & 0.55\,\%   \\
\hline
\end{tabular}
\end{table}

The actual performance of the designs generated are compared with the targets for the testing dataset, as shown in Fig.~\ref{fig:design_accuracy_plot}. The generated values are evaluated using the meanline model as introduced in Section~\ref{subsubsection:single_target_generation}. The line of $y=x$ is plotted for reference, representing the line of perfect generation with 0 \,\% $\textrm{RE}$ from Eq.~\eqref{equation:error}. Greater deviation of the scattered points implies less accurate designs. The corresponding distribution are also plotted for reference. It is clear from Fig.~\ref{fig:design_accuracy_plot} that the overall agreement of the generated vs. targeted values of $\textrm{PR}$ and $\eta$ are reported for both 3D and 1D model. Both models perform better on pressure ratio than on efficiency, which is likely a result from the skewed $\eta$ distribution in the training dataset shown in Fig.~\ref{fig:compressor_performance_distribution}.

To further quantify the model design accuracy, metrics outlined in Section~\ref{subsubsection:design_accuracy} are summarised in Table~\ref{table:design_accuracy_table} for the testing dataset.
It can be seen that the two diffusion models indicate comparable performance in the single-target generation tasks, with the 1D diffusion model performing slightly better than the 3D model across all metrics. This is because the 1D diffusion model has a much lower data dimension with a simpler model deployment. In contrast, the 3D model deployment is more complicated due to the dual diffusion model architecture and the 3D-to-1D post-processing process, which adds additional uncertainties during accuracy evaluation. Nevertheless, the 3D diffusion model still shows promising results with high design solution accuracies, proving the concept of directly applying the diffusion model to 3D geometry data. This paves the way towards more complicated blade geometries in the future, such as nature-inspired design, which fundamentally does not depend on geometry parametrisation~\cite{Juangphanich2019}.

\subsection{Design Variety Results}
\begin{figure}
    \centering
    \begin{subfigure}{0.49\textwidth}
        \centering
        \includegraphics[width=0.9\textwidth]{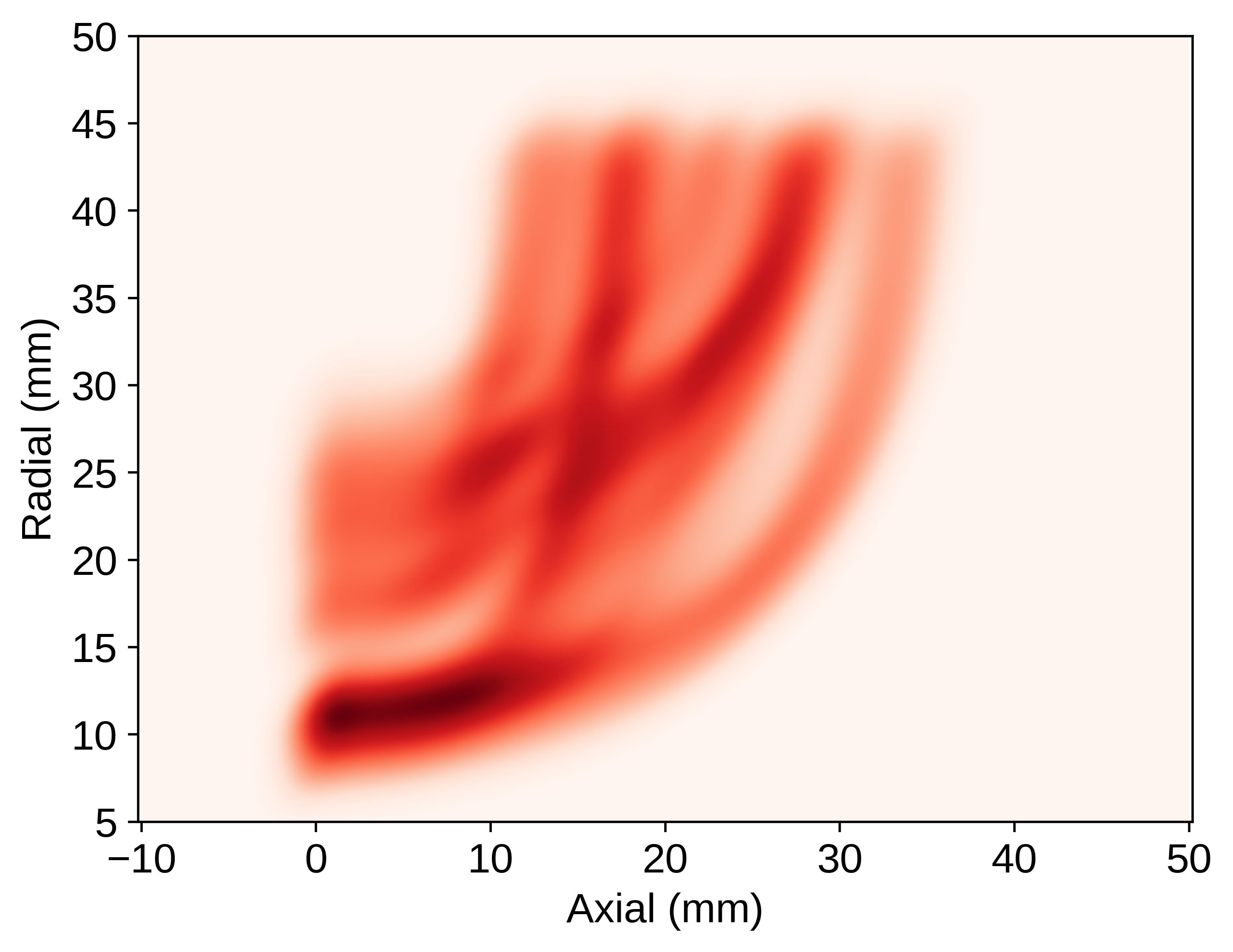}
        \caption{Solution Set $\mathcal{A}_{\text{3D}}$}
        \label{subfig:3D_model_generated_distribution}
    \end{subfigure} 
    \begin{subfigure}{0.49\textwidth}
        \centering
        \includegraphics[width=0.9\textwidth]{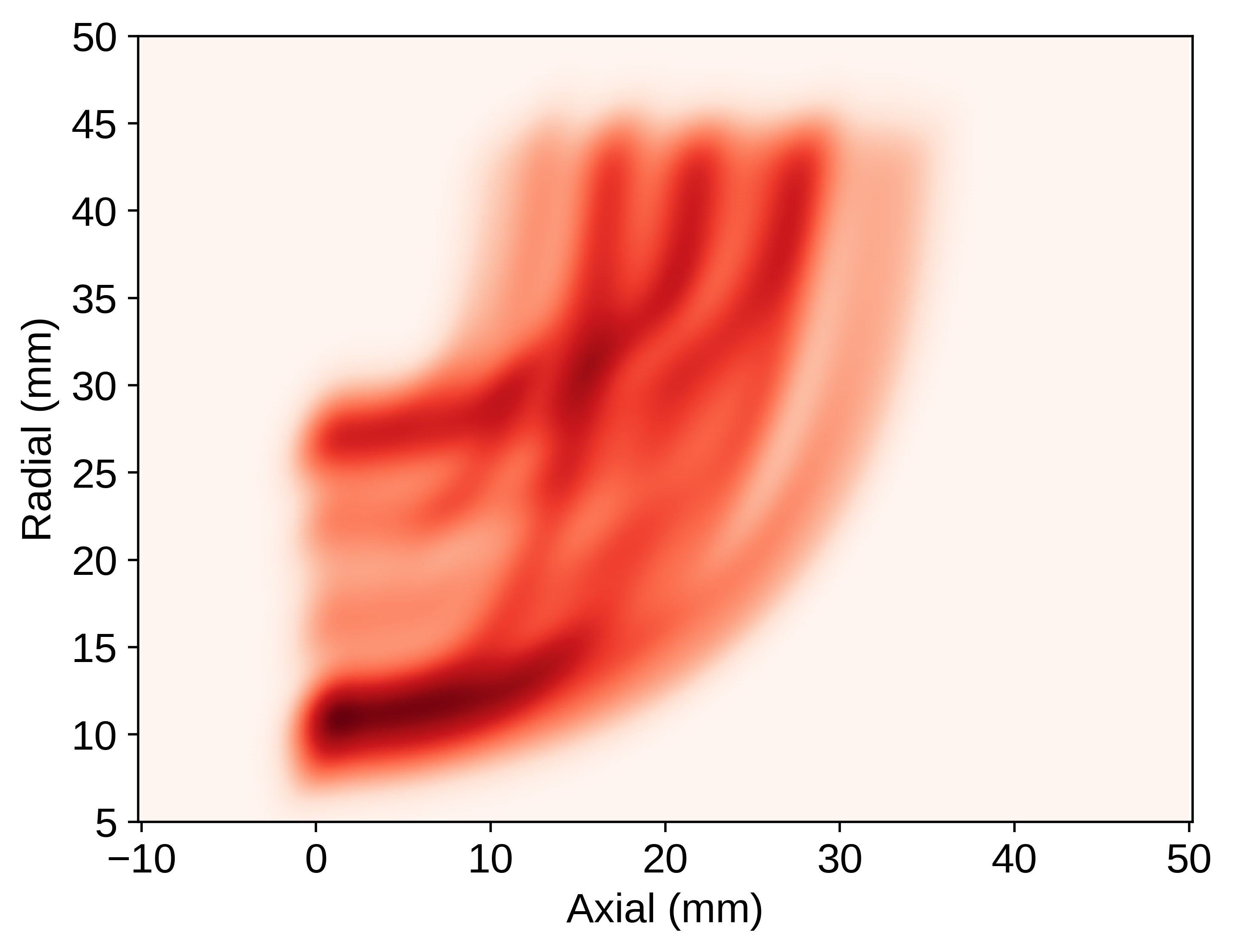}
        \caption{Solution Set $\mathcal{A}_{\text{1D}}$}
        \label{subfig:1D_model_generated_distribution}
    \end{subfigure}
    \begin{subfigure}{0.49\textwidth}
        \centering
        \includegraphics[width=0.9\textwidth]{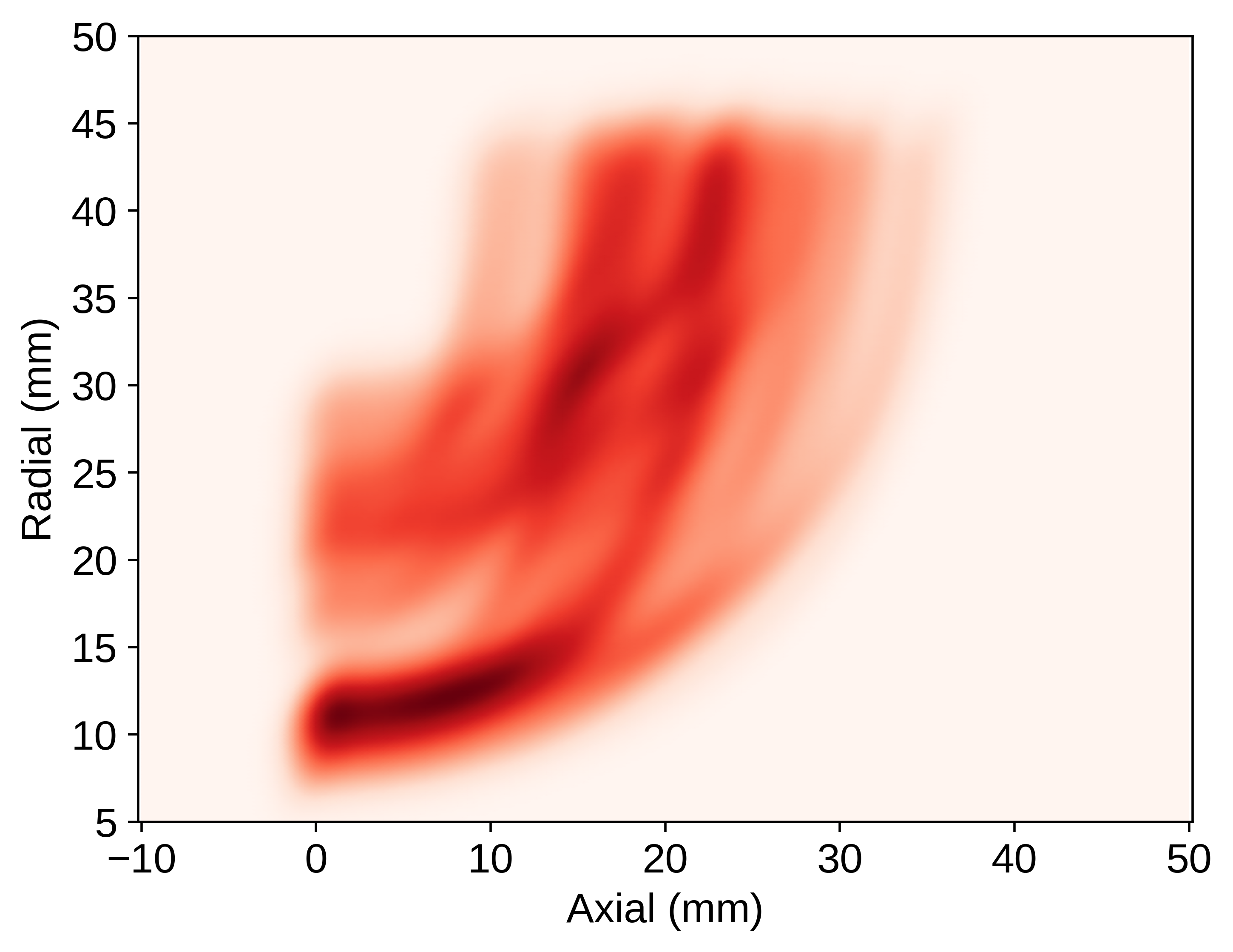}
        \caption{Solution Set $\mathcal{B}_{1}$}
        \label{subfig:meanline_model_generated_distribution_1}
    \end{subfigure}
    \begin{subfigure}{0.49\textwidth}
        \centering
        \includegraphics[width=0.9\textwidth]{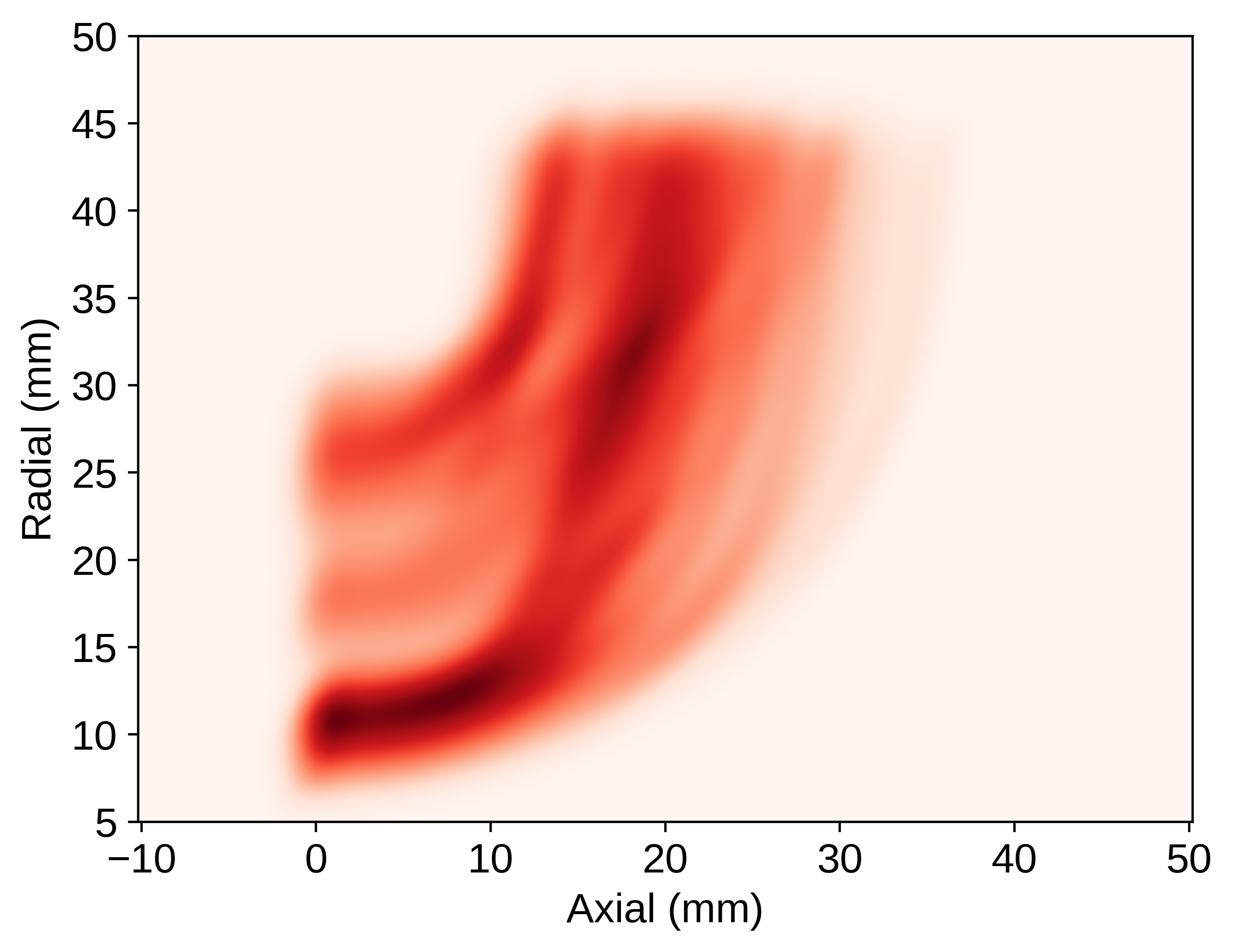}
        \caption{Solution Set $\mathcal{B}_{2}$}
        \label{subfig:meanline_model_generated_distribution_2}
    \end{subfigure}
    \caption{The KDE Plots of the Four Solution Sets on the Meridional Plane}
    \label{fig:solution_sets}
\end{figure}

\begin{figure}
    \centering
    \begin{subfigure}{\textwidth}
        \centering
        \includegraphics[width=\textwidth]{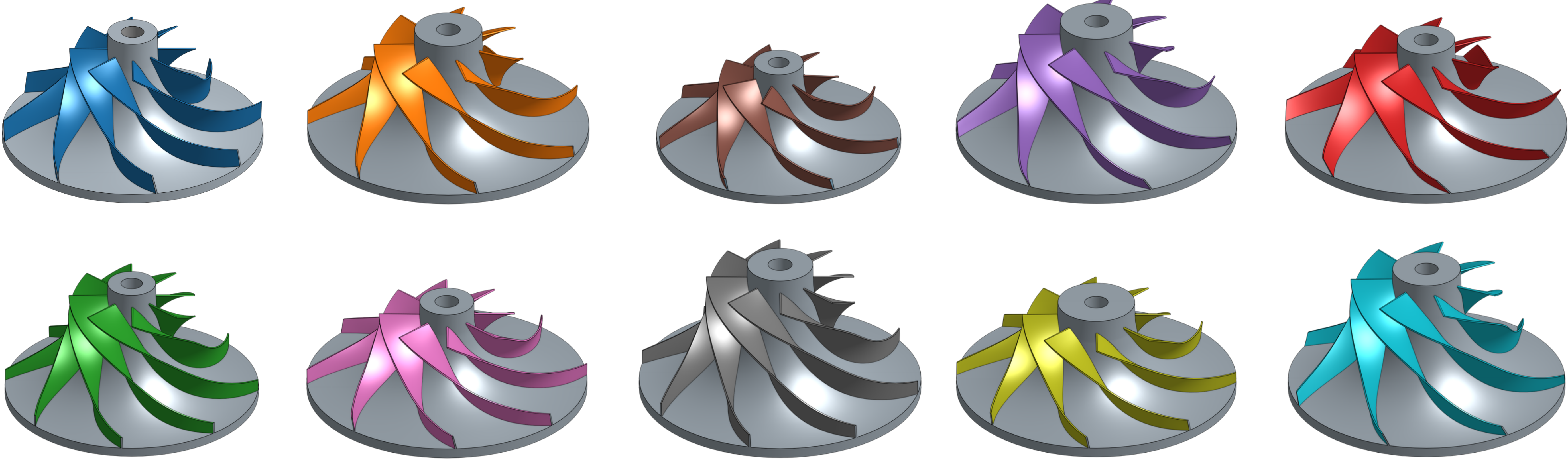}
        \caption{Design Pool}
        \label{subfig:design_pool}
    \end{subfigure} 
        \hfill
    \begin{subfigure}{0.47\textwidth}
        \centering
        \includegraphics[width=\textwidth]{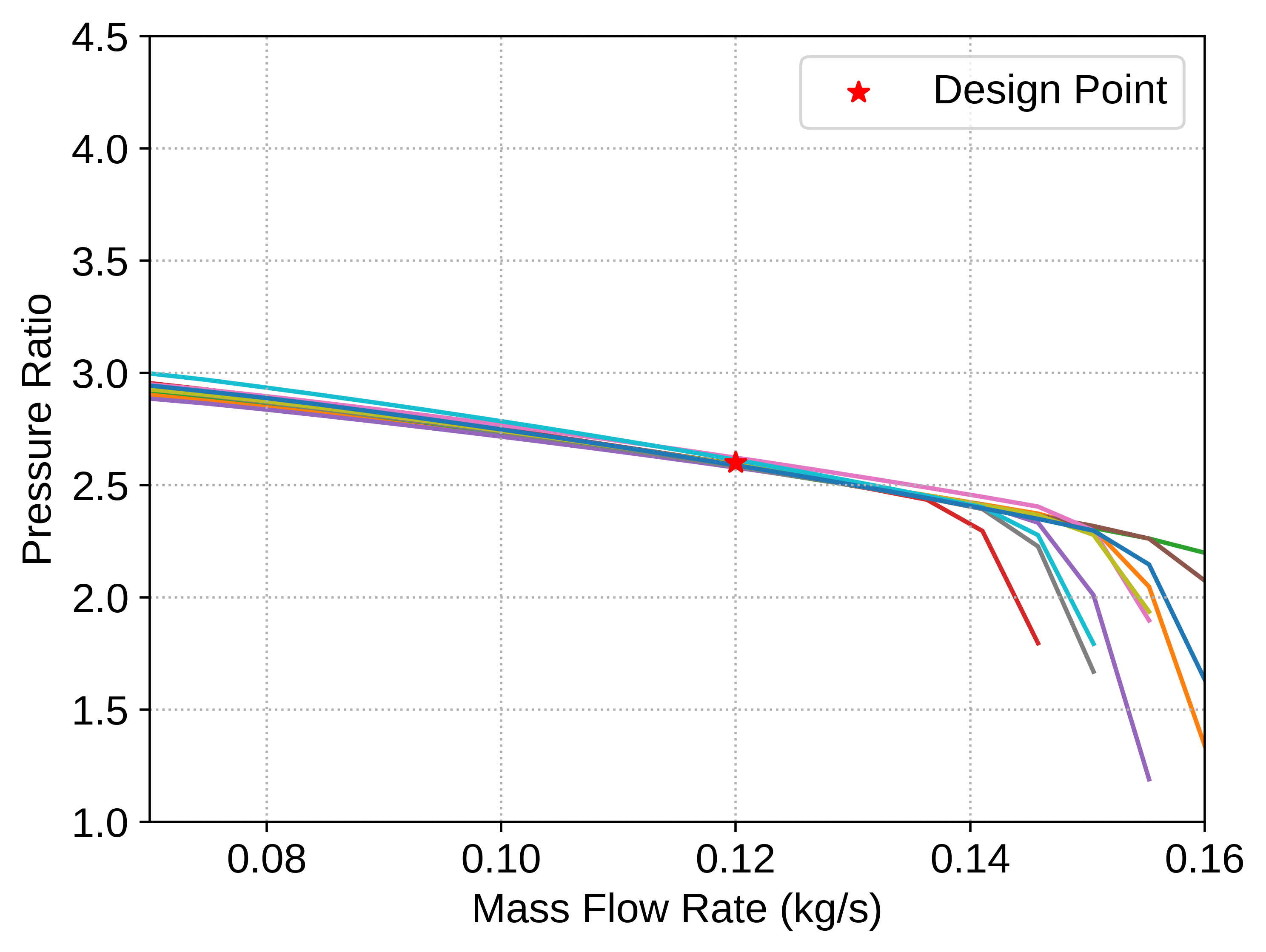}
        \caption{$\textrm{PR}$ at off-design $\dot{m}$}
        \label{subfig:pr_off_design}
    \end{subfigure}
        \hfill
    \begin{subfigure}{0.49\textwidth}
        \centering
        \includegraphics[width=\textwidth]{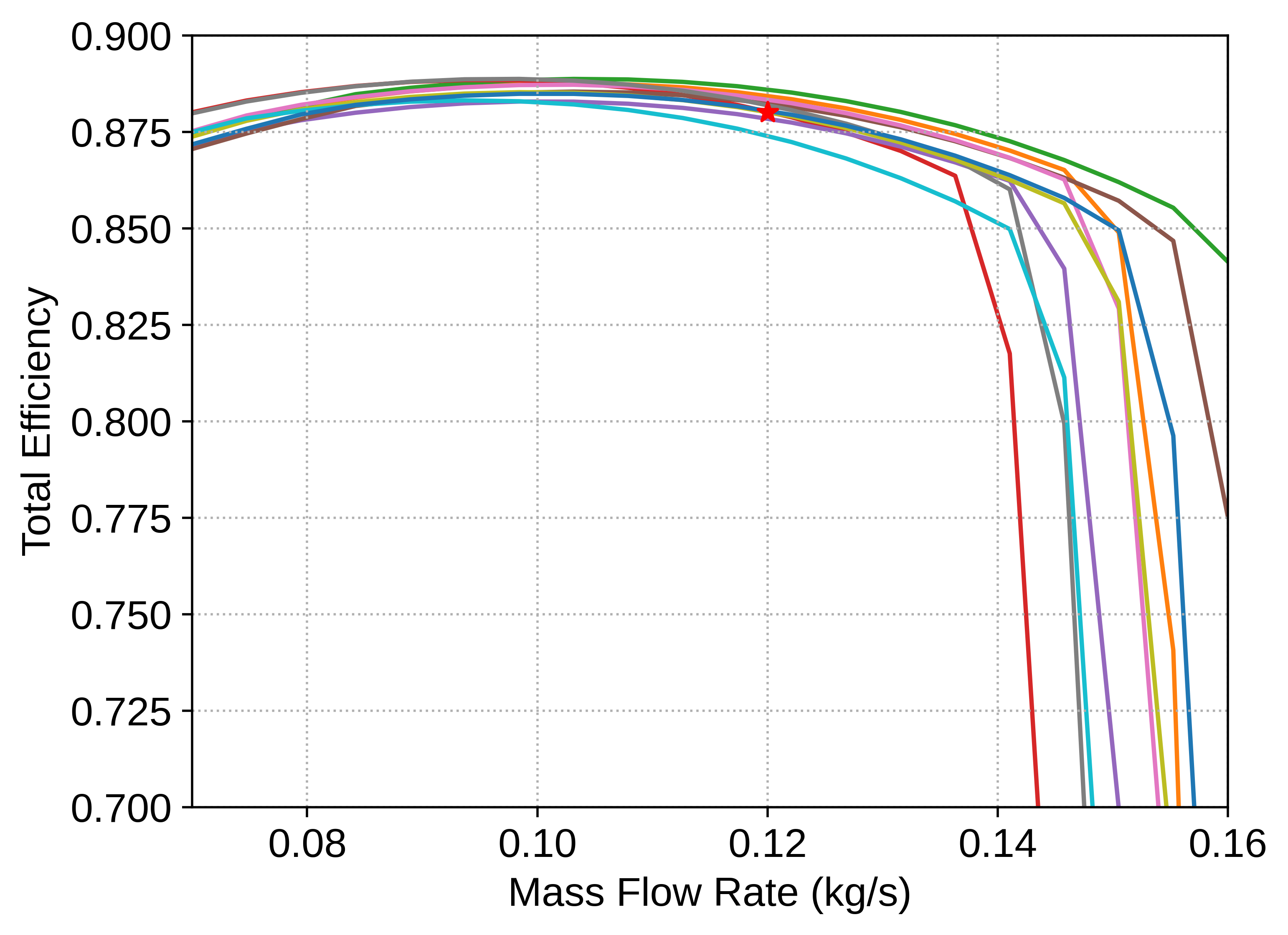}
        \caption{$\eta$ at off-design $\dot{m}$}
        \label{subfig:eta_off_design}
    \end{subfigure}
    
    \caption{Multi-Targets Generation Results}
    \label{fig:multi_targets_generation_results}
\end{figure}

Following the methods introduced in Section~\ref{subsubsection:desisgn_diversity}, both the two diffusion models (1D and 3D) are tasked to generate 100 design solutions corresponds to the target operating conditions of $\dot{m} = 0.09 ~\textrm{kg\,s}^{-1}, \omega = 80000 ~\textrm{RPM},$ and desired performance of $\textrm{PR} = 2.70, \eta = 0.85$. 
% For clarity, these diffusion model generated solutions sets are denoted by $\mathcal{A}_{\text{1D}}$ and $\mathcal{A}_{\text{3D}}$ in this section. The other two solutions sets that are obtained from direct design parameter sampling are denoted by Sets $\mathcal{B}_{1}$ and $\mathcal{B}_{2}$. 
Figure~\ref{fig:solution_sets} illustrates the kernel density estimate of the four solution sets on the meridional ($x$-$r$) plane. The corresponding distribution similarity metrics are listed in Table~\ref{table:design_variety_table}. 

It can be seen that both the 1D and 3D diffusion models provide very diverse design solutions (sets $\mathcal{A}_{\text{1D}}$ and $\mathcal{A}_{\text{3D}}$) compared to the solutions directly sampled from design parameters ($\mathcal{B}_{1}$). The values of similarity metrics between Sets $\mathcal{A}_{\text{1D}}$, $\mathcal{A}_{\text{3D}}$ and Set $\mathcal{B}_{1}$ are very close to the values between Sets $\mathcal{B}_{1}$ and $\mathcal{B}_{2}$. Overall, the 3D diffusion model provides marginally better matching of the distribution than the 1D model. 
%The values of similarity metrics between , indicating that by proving more degrees of freedom of design, the 3D diffusion model can explore more effectively the entire design space. In contrast, since the 1D diffusion model is trained directly using parametrised geometries, its ability to design variety is weaker, leading to less effective exploration. 

\begin{table}
    \centering
    \caption{Design Variety Metrics}
    \label{table:design_variety_table}
    \begin{tabular}{lccc}
    \hline
    \textbf{Pair} & \textbf{CD ($\downarrow$ better)} &\textbf{SSIM ($\uparrow$ better)} & \textbf{JSD ($\downarrow$ better)} \\
    \hline
    $\mathcal{A}_{\text{3D}}-\mathcal{B}_{1}$ & 2.31 mm & 0.970 & 0.062 \\
    $\mathcal{A}_{\text{1D}}-\mathcal{B}_{1}$ & 2.32 mm & 0.971 & 0.050 \\
    $\mathcal{B}_{1}-\mathcal{B}_{2}$ & 2.17 mm & 0.973 & 0.049 \\
    \hline
    \end{tabular}
\end{table}

\subsection{Multi-target Generation Results}
\label{subsection:multi_target_generation_results}
Figure~\ref{subfig:design_pool} shows an example of the non-deterministic mapping using the 3D diffusion model, where a design pool containing ten different geometries is generated. All of the geometries shown is generated with following target operating consitions of $\dot{m} = 0.12 ~\textrm{kg\,s}^{-1}, \omega = 90000 ~\textrm{RPM},$ and desired performance of $\textrm{PR} = 2.60, \eta = 0.88$. The corresponding actual value calculated from meanline model after post-processing have $\textrm{RE}<\varepsilon$. A wide variety of designs are visible from Fig.~\ref{subfig:design_pool}, showing the potential of the model for multi-target generation. 

Following the procedure outlined in Fig.~\ref{fig:generation_flow_charts}, the designs can be further filtered and selected using additional targets, and here we use the off-design performance (i.e., at mass flow rates other than the designed values) as an example. Using the meanline model from Section~\ref{section: Performance Database Generation}, the compressor maps for the candidates in the design pool can be calculated at the same rotational speed $\omega$ but at different mass flow rates $\dot{m}$. The performance maps of total stage pressure ratio and efficiency are shown in Figs.~\ref{subfig:pr_off_design} and \ref{subfig:eta_off_design}, respectively, with the red star marking the desired performance. It can be clearly seen that, although the ten designs can achieve the design point performance within the tolerance ($\textrm{RE} < 1\,\%$), they exhibit significantly different off-design performances, and the best-performing profile can be selected according to the particular application requirement, such as maximising operating range. In fact, once trained, this selection can be flexibly extended to any targets, both geometrically and aerodynamically, without the need to retrain the diffusion model.

\subsection{3D CFD Analysis}

\begin{figure*}
    \centering
    \includegraphics[width=\textwidth]{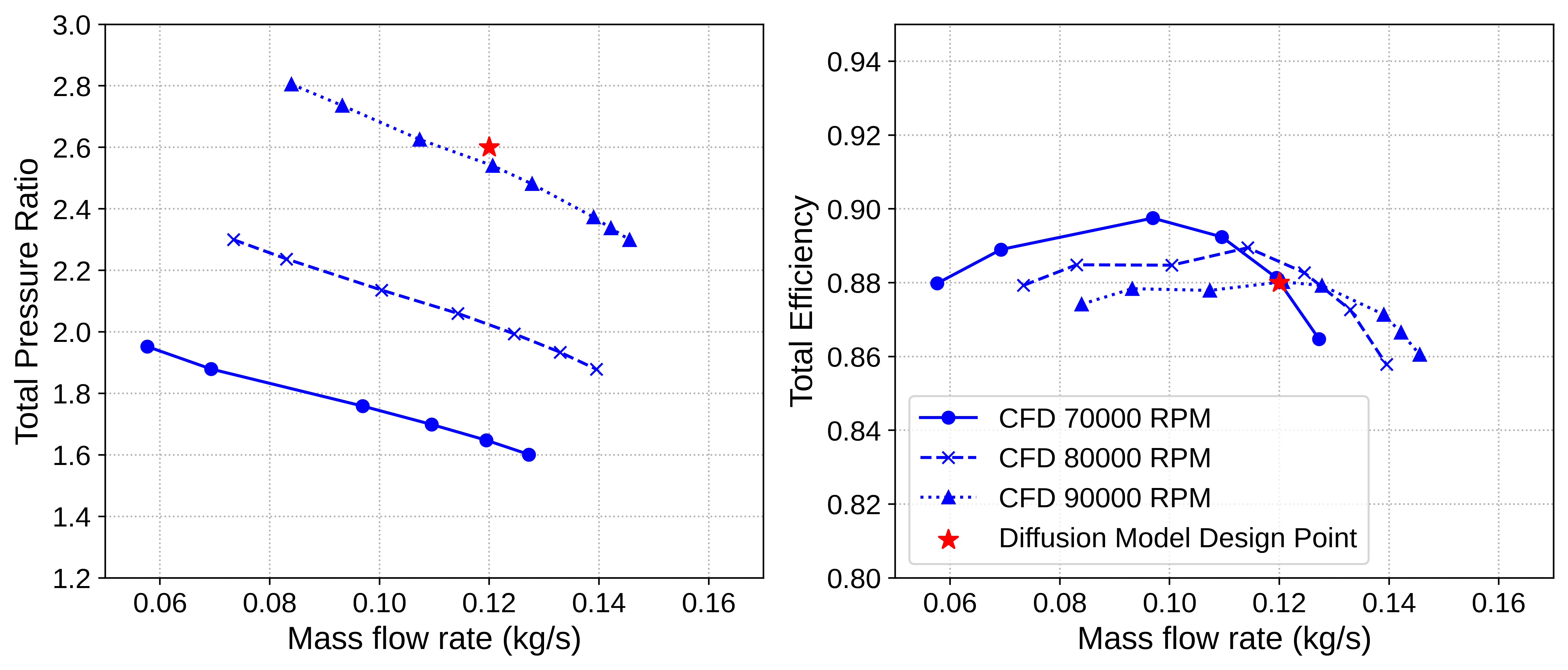}
    \caption{Compressor maps produced from the CFD simulations}
    \label{fig:diffusion_model_geometry_cfd}
\end{figure*}

In order to assess the physical feasibility of the generated designs, detailed CFD analysis is conducted on one compressor geometry generated from the diffusion model (the blue-colored geometry in Fig.\ref{subfig:design_pool}), with design pressure ratio 2.6 and efficiency 0.88, at operating point of 90000 RPM and $0.12 \textrm{ kg\,s}^{-1}$. CFD simulations were conducted using \textit{ANSYS CFX} with the $k$-$\omega$ SST turbulence model, which has been extensively validated for turbomachinery applications~\cite{Mentersst1994}. The inlet section is defined using stagnation properties (total pressure and total temperature) with a flow direction normal to the inlet boundary. A static pressure boundary condition was applied at the outlet to improve simulation convergence.The mesh was generated using \textit{ANSYS TurboGrid}, employing structured hexahedral topology to ensure high-quality elements throughout the domain. Boundary layer resolution was maintained at $y^+ < 2$ throughout all blade surfaces to ensure accurate capture of near-wall flow physics. In total, the computational domain consists of approximately 4 million cells. The numerical setup was selected based on previous experience with CFD simulations of centrifugal compressors of comparable scale and operating conditions~\cite{Papachristodoulou2026}.

Simulations were performed for three rotational speeds (70000, 80000, 90000 RPM) and cover the mass flow rate in the range $[0.06, 0.16] \textrm{ kg\,s}^{-1}$. The compressor map post-processed from the CFD results is shown in Fig.~\ref{fig:diffusion_model_geometry_cfd} with a red star marking the 3D diffusion model's specified target performance ($\textrm{PR}$ and $\eta$) as the design point. Compared to the target performances specified to the diffusion model, the actual performance of the generated compressor deviates by 2\,\% for pressure ratio and by 0.2\,\% for efficiency. This performance deviation is within expectation and is considered acceptable, given the inherent uncertainties of the diffusion and meanline models. Overall, the CFD analysis validates the quality and physical validity of the compressor geometry generated through the framework presented in the current study.

\section{Conclusion and Future Scope}
\label{section:conclusion}

In summary, this paper presents a diffusion model-powered 3D turbomachinery inverse design framework, and we thoroughly assessed its design capability using centrifugal compressor as an example. The model achieved accurate, diverse, and constraint-satisfying geometry generations across broad ranges of operating conditions, showing the potential of a new paradigm of using generative models for turbomachinery designs. The diffusion model can provide not only accurate but also diverse solutions, offering a new efficient and flexible inverse design paradigm in turbomachinery. This paper also presents the first diffusion model that is directly trained on 3D turbomachinery geometries, paving the way towards more advanced design and optimisation methods. 

As a concluding remark, potential future works are identified as follows:
\begin{itemize}
    \item For demonstration purposes, the dataset used in this paper is retrieved using parametrised geometries and meanline models (as explained in Section~\ref{subsection:data_curation_and_processing}). This paper reveals the potential of the proposed design framework and can be applied in the future to datasets with higher complexity and simulation fidelity. 
    \item As a proof-of-concept study, this study mainly demonstrates the methodology using compressor impeller geometries. Additional components, such as impeller splitter blades, diffusers, volutes and recirculation channels, may be considered and incorporated into the framework, hence achieving holistic generative turbomachinery system design.
    \item The ordered point coordinate data structure is opted in this paper as an effective direct geometry encoding method, which demonstrates excellent diffusion model training outcomes. As future work, additional geometry encoding approaches like point clouds~\cite{Luo2021}, voxel grids~\cite{Ren2024}, and 3D signed distance functions (SDF)~\cite{Park2025} can be tested for greater design flexibility.
\end{itemize}


% ===============================================================================
% ===============================================================================

\printcredits

\bibliographystyle{elsarticle-num-names}
% \bibliographystyle{cas-model2-names}

\bibliography{cas-refs}

\end{document}